\documentclass[submit]{smj}
\usepackage[utf8]{inputenc}

\RequirePackage{etex}

\usepackage{xfrac}
\usepackage{pdfpages}
\usepackage{siunitx}
\usepackage{enumitem}
\usepackage[export]{adjustbox}
\usepackage{needspace}
\usepackage{tabularx}
\usepackage{wrapfig}
\usepackage{pdflscape}
\usepackage[referable]{threeparttablex}
\usepackage{mathptmx}
\usepackage{titlesec}
\usepackage{autonum}
\usepackage{setspace}

\titleformat{\section}
  {\normalfont\Large\bfseries}{\thesection}{1em}{}

\usepackage[labelfont = bf, justification = justified, singlelinecheck = false, figurename = Figure, tablename = Table]{caption}
\usepackage{bm}
\usepackage{soul}

\usepackage[titletoc,toc,title]{appendix}
\usepackage[position=top,singlelinecheck=off]{subfig}

\usepackage{longtable}
\usepackage{multirow}
\usepackage{booktabs} 
\usepackage{afterpage}
\usepackage{relsize}
\usepackage{cprotect}
\usepackage{fancyhdr}
\newcounter{tablenote}[table]

\usepackage{lipsum}
\usepackage{microtype}

\usepackage{array}
\newcolumntype{P}[1]{>{\centering\arraybackslash}p{#1}}
\allowdisplaybreaks

\fancypagestyle{landscape}{
  \fancyhf{}
  \fancyfoot[C]{\thepage}
  \renewcommand{\headrulewidth}{0pt}
  
}

\Author{Divan A. Burger\Affil{1}\Affil{2}\ORCID{0000-0001-8096-6371}, 
        Janet van Niekerk\Affil{3}\ORCID{0000-0002-4334-2057}, \and Emmanuel Lesaffre\Affil{4}\Affil{5}\ORCID{0000-0002-3747-6905}
}
\AuthorRunning{}

\Affiliations{
    \item Cytel Inc., 1050 Winter Street, Waltham, 02451, MA, USA
    \item Department of Mathematical Statistics and Actuarial Science, University of the Free State, Bloemfontein, South Africa
    \item Department of Statistics, University of Pretoria, Pretoria, South Africa
    \item I-BioStat, KU Leuven, Leuven, Belgium
    \item Department of Statistics and Actuarial Science, Stellenbosch University, South Africa
}

\CorrAddress{Divan Burger, Cytel Inc., 1050 Winter Street, Waltham, 02451, MA, USA}
\CorrEmail{divanaburger@gmail.com}
\CorrPhone{(+99)\;99\;999\;9999}

\Title{Robust median regression for count data with general lower truncation using a contaminated discrete Weibull model}
\TitleRunning{}

\begin{document}

    \maketitle

    \newpage

    {\bf Abstract:} In right-skewed count data, the mean is disproportionately affected by a long upper tail, whereas the median remains a more representative measure of central tendency. Discrete Weibull (DW) regression links covariates to a shifted median, which in turn induces the exact integer median; however, a single DW component can fit poorly when the observed count distribution has a markedly heavier upper tail than a single-component model can accommodate. We propose a contaminated DW (cDW) regression that augments the baseline DW distribution with a more dispersed secondary component within a finite mixture while retaining a single shifted-median link. This mixture accommodates extreme counts more effectively, thereby stabilizing the median-based regression coefficients. The model accommodates general lower truncation at an arbitrary threshold $c$, including $c = 1$ for strictly positive outcomes and $c = 0$ for nonnegative counts, and is estimated using a straightforward Bayesian Markov chain Monte Carlo algorithm implemented in JAGS; R code accompanies the paper. Applied to hospital length-of-stay data, the cDW regression reduces the influence of outliers and achieves superior predictive performance relative to a single-component DW model, as demonstrated by leave-one-out cross-validation and a Kullback-Leibler influence diagnostic. Simulation experiments show that, under strongly heavy-tailed mixture settings, the cTDW model accurately recovers the regression coefficients and improves on the single-component TDW model. Because the added tail component can increase probability mass at both extremes, we further recommend embedding the cDW in a hurdle framework when structural zeros are present: the zero probability is modeled separately, and the heavy-tail mixture is applied only to positive counts. The cDW regression model provides a robust, median-centered alternative for analyzing skewed, possibly truncated count outcomes.


    \newpage
    
    \section{Introduction} \label{sec:intro}

    Count data arise frequently in fields such as epidemiology, healthcare, insurance, and sociology, where the outcome of interest is the number of events or occurrences. A classical modeling choice is the Poisson distribution, but overdispersion, where the variance exceeds the mean, often arises in practice. Many extensions have been proposed to handle overdispersion, including negative binomial (NB) models and zero-inflation/hurdle approaches when there are excess zeros \citep{hilbe2011negative}.
    
    A separate driver of overdispersion is the presence of outliers or heavy tails, for which standard mean-based approaches (e.g., the NB) may not adequately accommodate extreme observations. In some settings, the empirical distribution exhibits a substantially heavier upper tail than a single-component count model can capture. If left unmodeled, such tail behavior may disproportionately influence estimates of the mean or regression coefficients, thereby negatively impacting inference and prediction.
    
    Instead of focusing on the mean, which can be strongly affected by outliers, one can target the median, often a more robust measure of central tendency for skewed or heavy-tailed counts. The discrete Weibull (DW) distribution introduced by Nakagawa \& Osaki \citep{nakagawa1975discrete} was extended by Kalktawi \citep{kalktawi2017} and Burger et al. \citep{burger2020robust} to allow for covariates through a \emph{shifted median}, producing a flexible approach for skewed count data. Still, even a median-based DW can fail to capture extremely large values when a subset of observations exhibits substantially heavier tails than the bulk.
    
    A natural strategy is to adopt a two-component mixture in which one component is more dispersed than the other while both share the same center. Such constructions are well-established for continuous data, such as contaminated normal distributions \citep{Hampel2011}. In the count-data setting, robust methods often focus on zero-inflated models, although contamination has also been introduced directly into count regression models, including contaminated NB mean regression \citep{otto2025contaminated} and contaminated beta-binomial regression for bounded counts \citep{otto2026modeling}. The contaminated beta-binomial model of Otto et al. \citep{otto2026modeling} is particularly relevant here because it also allows the mixture-weight parameter in a contaminated count model to range over the full open unit interval. In the present paper, we use the term ``contaminated'' in this broader robust-modeling sense, namely the addition of a heavier-tailed component to a baseline distribution, without requiring that the heavier-tailed component represent a small minority of observations. Indeed, while a heavier right tail is desirable for capturing extreme counts, the mixture can also inflate lower counts, including zeros, that may not truly exist. Such unintended zero inflation complicates model interpretation and parameter estimation. Although not solely for this reason, directly mixing a baseline count distribution with a heavier-tailed subcomponent remains relatively uncommon in routine applications.
    
    In this paper, we propose a contaminated DW framework that augments the DW distribution of Burger et al. \citep{BURGER2021} with a heavier-tailed subcomponent and focuses on regression through a shifted-median parameterization that preserves the exact integer median. We allow for general lower truncation at $c$; in applications, we focus on $c=1$ (strictly positive outcomes such as length-of-stay (LOS)), while setting $c=0$ recovers the usual (untruncated) DW for nonnegative counts. Our approach extends naturally to regression by linking covariates to the shifted median parameter. The mixture weights determine how probability mass is split between the narrower and heavier-tailed subcomponents.
    
    We first establish the theoretical structure of the truncated DW (TDW) distribution, deriving raw-moment expansions to illustrate its tail limitations. We then introduce the contaminated TDW (cTDW) distribution, which addresses these limitations by adding a heavier-tailed subcomponent, and we demonstrate how to perform Bayesian inference for both regression models. Using a hospital LOS dataset (with no zero-day stays), we show that moderate outliers can noticeably shift parameter estimates in a single-component TDW model, whereas the cTDW model remains more stable and yields improved model adequacy and predictive performance.
    
    Numerous robust methods for outlier-prone count data relate closely to our research, including robust M-estimation in generalized linear models \citep{cantoni2001robust} and mixture-based methods that explicitly model outliers \citep{beath2018mixture}. Recent Bayesian frameworks incorporate heavy-tailed priors \citep{datta2016bayesian} or simultaneously handle zero inflation and upper-tail outliers in count data~\citep{Hamura03012025}. However, many of these remain mean-based. Our framework aims directly at the exact integer median through a shifted-median parameterization.

    NB regression is mean-based, whereas under truncation, the target $E\left(X \left| X \ge c\right.\right)$ couples the mean with $P\left(X < c\right)$ and lacks a simple one-parameter link. In the cTDW, the shifted-median parameter $m^*$ remains in one-to-one correspondence with the exact integer median $m_{\mathrm{med}} = \left\lceil m^* - 1 \right\rceil$, thereby preserving direct median-based interpretation under truncation. A contaminated version of the truncated NB (TNB) would remain mean-centered and, under truncation, does not admit a simple one-parameter link that preserves the truncated mean; it also has no closed-form median. Accordingly, developing a contaminated TNB (cTNB) is beyond our scope. We include a TNB comparator as a predictive benchmark.
    
    The remainder of this paper is organized as follows. \autoref{sec:data} describes the Arizona LOS dataset that motivates our work. \autoref{sec:t_count_distr} details the TDW distribution and its contaminated version (cTDW). \autoref{sec:regression_models} explains how we incorporate covariates via a shifted median link. In \autoref{sec:model_adequacy}, we present simulation-based residual diagnostics and Kullback-Leibler (K-L) divergence checks to assess model adequacy and identify outliers, while \autoref{sec:model_comparison} describes model comparison via leave-one-out (LOO) cross-validation. \autoref{sec:application} applies the TDW and cTDW models to the hospital LOS data. \autoref{sec:simulation} reports simulation studies to compare each model's performance and robustness. Finally, \autoref{sec:conclusions} discusses the results and outlines a possible hurdle extension, although we do not undertake direct comparisons with other robust count methods here.

    \section{Arizona hospital length-of-stay dataset} \label{sec:data}

    Our primary dataset comes from a 1991 Arizona study examining hospital LOS among cardiac patients who underwent one of two revascularization procedures: coronary artery bypass graft (CABG) or percutaneous transluminal coronary angioplasty (PTCA) \citep{ArizonaMedpar1991}. CABG bypasses blocked or diseased arteries by grafting new vessels around them, while PTCA uses a balloon to dilate those arteries. Because CABG is typically more invasive, it often leads to longer recovery and thus longer LOS; PTCA is relatively less invasive with shorter recovery \citep{Fihn2012}. Accordingly, one might expect distinct LOS and risk profiles between these two procedures.
    
    The dataset records total hospital days along with admission type (elective vs. urgent/\allowbreak emergent) and patient sex (male vs. female). Analyzing these variables provides insights into how procedural choice, admission type, and sex influence LOS, which is useful for hospital resource planning and risk management. This dataset was featured by Hilbe \citep{hilbe2011negative} in his work on count data regression and is available in the \texttt{COUNT} package for R \citep{RCOUNT2025}.
    
    In this dataset, LOS is strictly positive by design. Let $Y_i$ be the LOS (days) for individual $i$; we therefore model $Y_i \in \left\{1,2,\ldots\right\}$ using the zero-truncated likelihood at $c=1$. This does not assume that $Y_i=0$ is possible; it encodes the known lower bound. Recoding $Z_i=Y_i-1$ and fitting an untruncated model are not, in general, equivalent. Truncation introduces a parameter-dependent normalizing constant and alters the estimand, except in special cases where the distribution admits a simple shift (e.g., the geometric). Thus, the interpretation of the original LOS scale is preserved only under the truncated specification.
    
\begin{figure}[ht]
        \centering
        \includegraphics[width=1.0\textwidth]{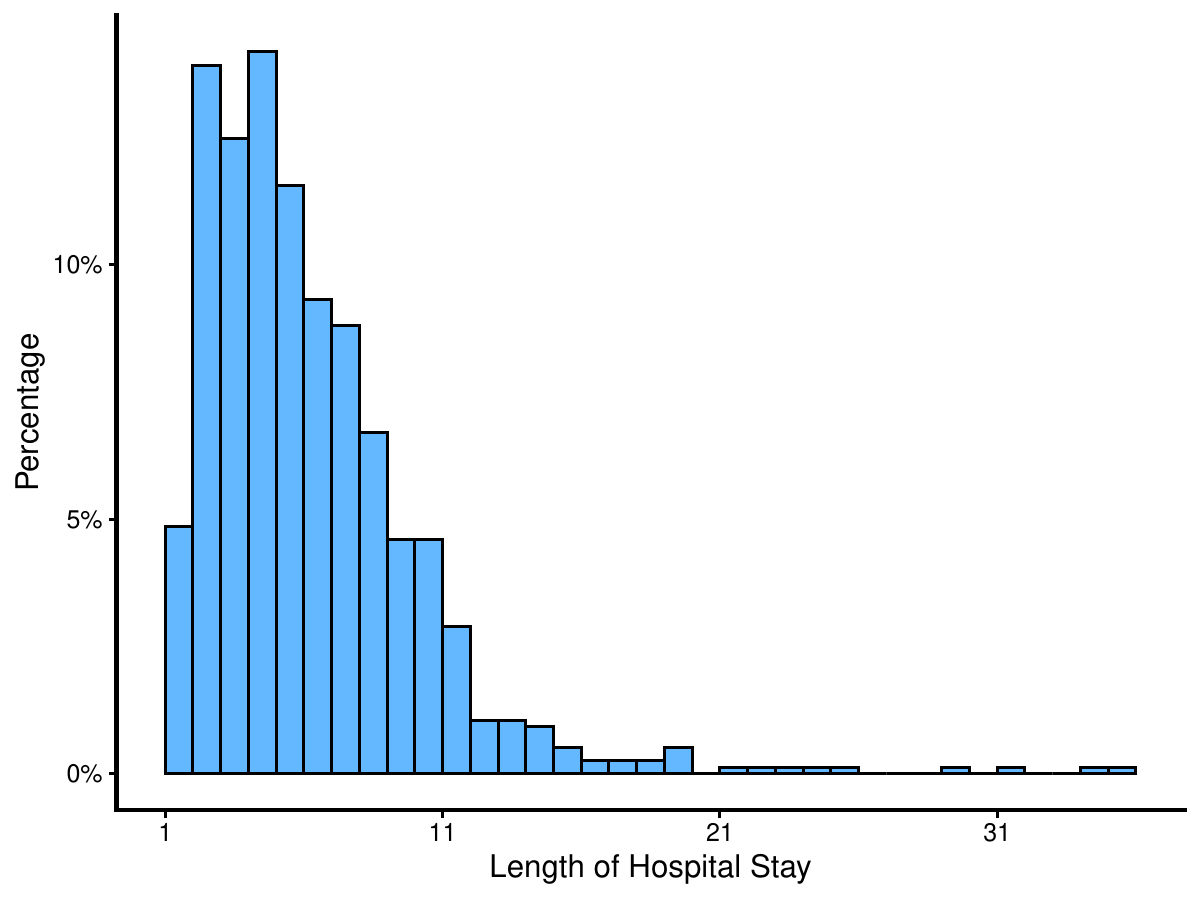}
        \caption{Distribution of length of hospital stay for male PTCA patients with urgent/emergent admissions.}
        \label{fig:LOS_histogram}
    \end{figure}

    \autoref{fig:LOS_histogram} displays the LOS distribution for male patients receiving PTCA on an urgent or emergent basis. The histogram is right-skewed with a pronounced heavy tail: although most stays last only a few days, some extend well beyond two weeks. Such skewness is common in LOS data and calls for models that accommodate heavy tails and outliers.

    For context, this Arizona LOS subset has often been analyzed with TNB models in the literature and teaching resources \citep{hilbe2011negative, RCOUNT2025, hardin2015regression}. To mirror that baseline and provide a mean-centered comparator, we also fit a TNB model and report the LOO information criterion (LOOIC). These alternatives target the mean of the untruncated distribution and, under truncation, do not yield a simple regression link to the truncated mean; by contrast, our TDW/cTDW framework links covariates directly to the smooth shifted-median parameter, which remains in one-to-one correspondence with the exact integer median after truncation and is robust to the heavy upper tail observed in \autoref{fig:LOS_histogram}.

    \section{Truncated count distributions} \label{sec:t_count_distr}

    We first introduce the TDW distribution (\autoref{sec:TDW_dist}), describe its probability mass function (PMF) and cumulative distribution function (CDF), and examine its Pearson kurtosis. Analyzing the kurtosis shows that while the single-component TDW distribution is flexible, it may still exhibit limited tail behavior, thereby motivating the mixture approach (cTDW) in \autoref{sec:contam_TDW}.

    \subsection{Truncated discrete Weibull distribution} \label{sec:TDW_dist}

    For the TDW distribution, we adopt the DW distribution of Nakagawa \& Osaki \citep{nakagawa1975discrete}, truncating at $c$ and renormalizing. Based on the standard binomial identity, our moment formulas employ a telescoping approach (see Rosen \citep{rosen2019discrete}).
    
    Let $X$ be a DW random variable on $\left\{0,1,2,\ldots\right\}$ with parameters $0 < q < 1$ and $\rho > 0$. Its PMF is
    \begin{equation}
        P_\text{DW}\left(X = n\right)
        = q^{n^\rho} - q^{\left(n+1\right)^\rho},
        \quad n = 0,1,2,\ldots
    \end{equation}
    Truncating at $c$ means defining
    \begin{equation}
        Y = X \left|\left(X \ge c\right)\right.,
    \end{equation}
    so the total probability from $n=0$ to $n=c-1$ is removed, and the remainder is renormalized by $1/q^{c^\rho}$. Thus, the PMF of the TDW distribution can be written as
    \begin{equation}
        P_\text{TDW}\left(Y = y\right)
        = \frac{q^{y^\rho} - q^{\left(y+1\right)^\rho}}{q^{c^\rho}},
        \quad y = c,c+1,\ldots ,
    \end{equation}
    and the CDF is
    \begin{equation}
        F_\text{TDW}\left(y\right)
        = 1 - q^{\left(\left(y+1\right)^\rho - c^\rho\right)},
        \quad y = c,c+1,\ldots
    \end{equation}
    The exact integer median $m_{\mathrm{med}}$ is the smallest integer $m$ with $F\left(m\right)\ge 0.5$. Equivalently,
    \begin{equation}
        q^{\left(\left(m+1\right)^\rho - c^\rho\right)} \le 0.5
        \Longleftrightarrow
        \left(m+1\right)^\rho \ge c^\rho + \frac{\ln\left(0.5\right)}{\ln\left(q\right)},
    \end{equation}
    which leads to
    \begin{equation}
        m_{\mathrm{med}}
        = \left\lceil
        \left(
        c^\rho + \frac{\ln\left(0.5\right)}{\ln\left(q\right)}\right)^{\frac{1}{\rho}}
        - 1
        \right\rceil.
    \end{equation}
    For simplicity (e.g., in regression modeling), one may drop the ceiling and treat
    \begin{equation}
        \left(m+1\right)^\rho
        = c^\rho + \frac{\ln\left(0.5\right)}{\ln\left(q\right)}
    \end{equation}
    as a real solution.
    
    To ensure the real median remains above $c$ (i.e., above the lower bound of the truncated domain), one can define
    \begin{equation}
        m^* = m + 1 > c,
        \quad
        \left(m^*\right)^\rho
        = c^\rho + \frac{\ln\left(0.5\right)}{\ln\left(q\right)}.
    \end{equation}
    Some authors replace $\rho$ with $\alpha = 1/\rho$, so that $\left(m^*\right)^\rho = \left(m^*\right)^{1/\alpha}$. Then
    \begin{equation}\label{eq:q_mstar_alpha}
        \left(m^*\right)^{\frac{1}{\alpha}}
        = c^{\frac{1}{\alpha}}
        + \frac{\ln\left(0.5\right)}{\ln\left(q\right)}
        \quad\Longrightarrow\quad
        q
        = \exp\left[
          \frac{\ln\left(0.5\right)}{\left(m^*\right)^{\frac{1}{\alpha}} - c^{\frac{1}{\alpha}}}
       \right].
    \end{equation}
    Hence, the TDW distribution can equivalently be parameterized by $\left(m^*, \alpha\right)$, with $m^* > c$ and $\alpha > 0$. Under this parameterization, the TDW survival function is
    \begin{equation} \label{eq:TDW_survival_mstar_alpha}
        P\left(Y \ge y \left|m^*, \alpha, c\right.\right)
        = q^{\left(y^{1/\alpha} - c^{1/\alpha}\right)}
        = 0.5^{\frac{y^{1/\alpha} - c^{1/\alpha}}{\left(m^*\right)^{1/\alpha} - c^{1/\alpha}}},
        \quad y = c,c+1,\ldots .
    \end{equation}
    In particular, substituting $y = m^*$ into the exponent on the underlying continuous scale gives
    \begin{equation}
        \frac{\left(m^*\right)^{1/\alpha} - c^{1/\alpha}}{\left(m^*\right)^{1/\alpha} - c^{1/\alpha}} = 1,
    \end{equation}
    so the survival expression equals $0.5$ at $y = m^*$. Thus, $m^*$ is the model's smooth shifted-median parameter and induces the exact integer median $m_{\mathrm{med}} = \left\lceil m^* - 1 \right\rceil$, differing from that integer median by no more than a single unit. Because the TDW distribution is discrete, $m^*$ is not itself the exact observed-data median; rather, it is a smooth parameter that is in one-to-one correspondence with the exact integer median. Equation \eqref{eq:TDW_survival_mstar_alpha} also shows directly how $\alpha$ controls tail behavior and dispersion: larger $\alpha$ (equivalently, smaller $\rho$) yields a more slowly decaying right tail, whereas smaller $\alpha$ yields a more rapidly decaying one. When $\alpha = 1$, the TDW reduces to the truncated geometric distribution.

\begin{figure}[ht]
        \centering
        \includegraphics[width=1.0\textwidth]{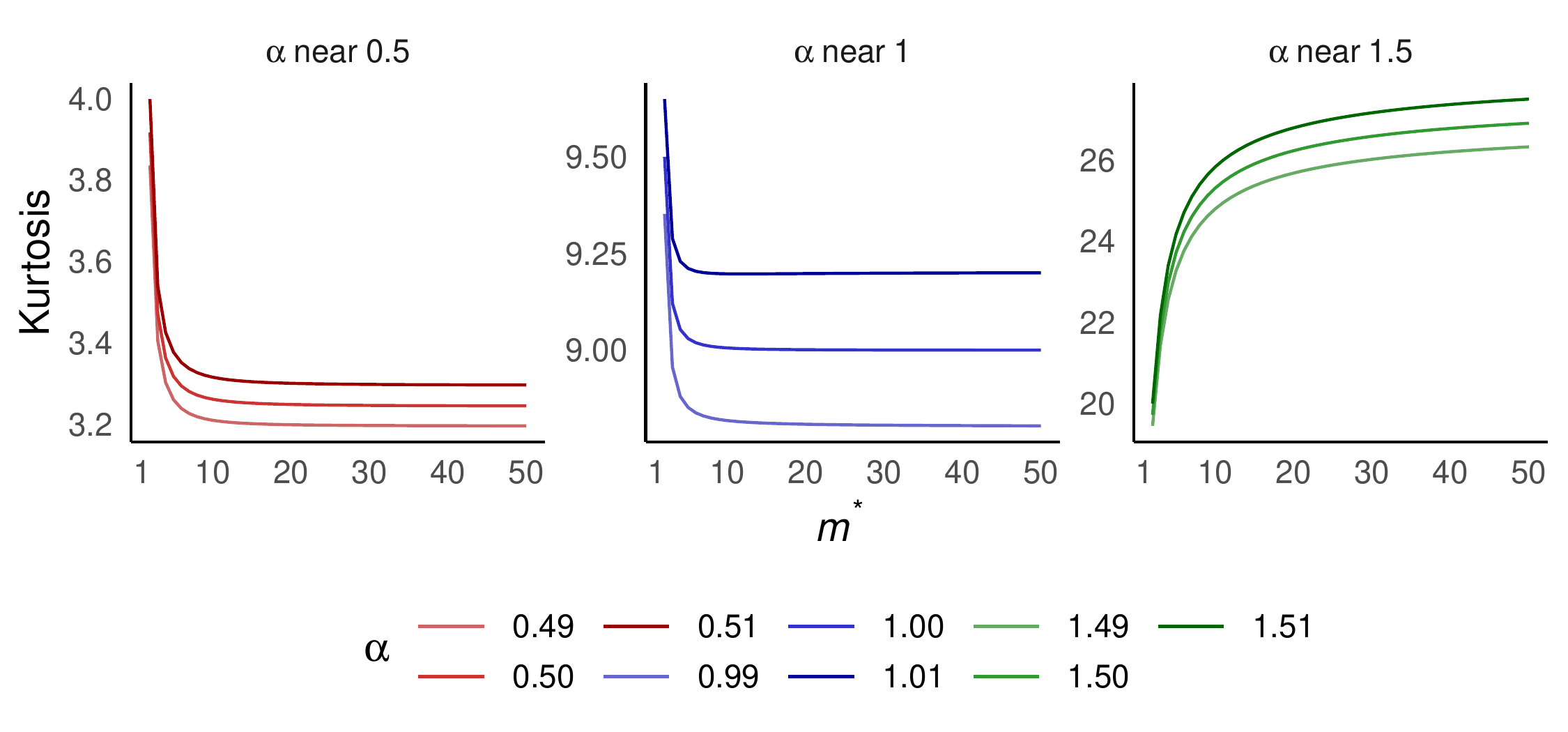}
        \caption{Pearson kurtosis of the TDW distribution plotted against the shifted median $m^*$. Each panel shows $\alpha$ near 0.5, 1, or 1.5 (with slight deviations shown as separate lines). When $\alpha=1$, the TDW distribution coincides with the truncated geometric distribution, whose kurtosis converges to 9. For $\alpha<1$, the TDW distribution has lighter tails than the geometric distribution and settles at a lower plateau. For $\alpha > 1$, the TDW distribution has heavier tails, resulting in a higher limiting kurtosis.}
        \label{fig:TDW_kurtosis}
    \end{figure}

    \autoref{sec:TDW_kurtosis} in \ref{sec:DW_dists_kurtosis} details the derivations of the raw moments and Pearson kurtosis for the TDW distribution. \autoref{fig:TDW_kurtosis} plots the kurtosis over various values of $m^*$ and $\alpha$ (with $c=1$), illustrating how the shifted median $m^*$ and the shape parameter $\alpha$ govern the distribution's tail behavior. When $\alpha=1$, the TDW coincides with a truncated geometric distribution on $\left\{c,c+1,\ldots\right\}$; substituting $\alpha=1$ into \eqref{eq:TDW_KURT} and using \eqref{eq:q_mstar_alpha} (so $m^*\to\infty$ implies $q\to 1$) shows that $\mathrm{Kurt}\left(Y\right)\to 9$. If $\alpha < 1$, the TDW distribution has lighter tails than this geometric baseline, and the kurtosis converges to a lower plateau (below 9) as $m^*$ grows. In contrast, if $\alpha > 1$, the TDW distribution has heavier tails than the geometric distribution, causing its kurtosis to settle at a higher limiting value (above 9).

    \subsection{Contaminated truncated discrete Weibull distribution} \label{sec:contam_TDW}
    
    As shown in \autoref{sec:TDW_dist}, the single-component TDW distribution may not adequately capture heavy-tail behavior or outliers in some applications. To address this, we consider a mixture of two TDW components that share the same smooth shifted-median parameter, $m^*$, and truncation, $c$, but differ in their dispersion parameters. Specifically, let one component have dispersion $\alpha$, while the other has dispersion $\eta \alpha$ with $\eta > 1$. We define the cTDW distribution:
    \begin{equation}
    \label{eq:CTDW_mixture_def}
        P_{\mathrm{cTDW}}\left(Y = y \left|m^*, \alpha, \eta, \delta, c\right.\right) = \delta P_{\mathrm{TDW}}\left(y \left|m^*, \alpha, c\right.\right) + \left(1 - \delta\right) P_{\mathrm{TDW}}\left(y \left|m^*, \eta \alpha, c\right.\right),
    \end{equation}
    for $y \in \left\{c, c + 1, \dots \right\}$, where $\delta \in \left(0, 1\right)$ is the mixture weight assigned to the narrower component and $\eta > 1$ controls tail inflation in the second component. Hence, with probability $\delta$, $Y$ is drawn from the narrower $\mathrm{TDW}\left(m^*, \alpha, c\right)$ component, whereas with probability $1 - \delta$, $Y$ is drawn from the more spread-out $\mathrm{TDW}\left(m^*, \eta \alpha, c\right)$ component. Several limiting cases are immediate. If $\delta = 1$, the cTDW reduces to $\mathrm{TDW}\left(m^*, \alpha, c\right)$, whereas if $\delta = 0$, it reduces to $\mathrm{TDW}\left(m^*, \eta\alpha, c\right)$. If $\eta = 1$, the two mixture components coincide, so the cTDW reduces to a single $\mathrm{TDW}\left(m^*, \alpha, c\right)$ regardless of $\delta$, and $\delta$ is unidentifiable. Allowing $\delta$ to lie in the full open unit interval is consistent with recent contaminated count-model formulations such as the contaminated beta-binomial model of Otto et al. \citep{otto2026modeling}, in which the corresponding mixture-weight parameter is also defined on $\left(0,1\right)$.

    Since both sub-distributions share the same pair $\left(m^*, c\right)$, they also imply the same exact integer median
    \begin{equation}
        m_{\mathrm{med}} = \left\lceil m^* - 1 \right\rceil.
    \end{equation}
    Let $F_1\left(y\right) = F_{\mathrm{TDW}}\left(y \left|m^*, \alpha, c\right.\right)$ and $F_2\left(y\right) = F_{\mathrm{TDW}}\left(y \left|m^*, \eta \alpha, c\right.\right)$ denote the two TDW CDFs. For completeness, we extend each CDF by defining $F_j\left(y\right) = 0$ for $y < c$, $j = 1,2$. Because the TDW distribution is discrete, one does not in general have $F_1\left(m_{\mathrm{med}}\right) = F_2\left(m_{\mathrm{med}}\right) = 0.5$. Instead, by the definition of the exact integer median,
    \begin{equation}
        F_j\left(m_{\mathrm{med}} - 1\right) < 0.5 \le F_j\left(m_{\mathrm{med}}\right), \quad j = 1,2.
    \end{equation}
    Now, define the mixture CDF
    \begin{equation}
        F_{\mathrm{mix}}\left(y\right) = \delta F_1\left(y\right) + \left(1 - \delta\right) F_2\left(y\right).
    \end{equation}
    Then
    \begin{equation}
        F_{\mathrm{mix}}\left(m_{\mathrm{med}} - 1\right)
        = \delta F_1\left(m_{\mathrm{med}} - 1\right) + \left(1 - \delta\right) F_2\left(m_{\mathrm{med}} - 1\right)
        < 0.5,
    \end{equation}
    and
    \begin{equation}
        F_{\mathrm{mix}}\left(m_{\mathrm{med}}\right)
        = \delta F_1\left(m_{\mathrm{med}}\right) + \left(1 - \delta\right) F_2\left(m_{\mathrm{med}}\right)
        \ge 0.5.
    \end{equation}
    Therefore,
    \begin{equation}
        F_{\mathrm{mix}}\left(m_{\mathrm{med}} - 1\right) < 0.5 \le F_{\mathrm{mix}}\left(m_{\mathrm{med}}\right),
    \end{equation}
    so $m_{\mathrm{med}}$ is also the exact integer median of the cTDW distribution.
    
\begin{figure}
        \centering
        \includegraphics[width=0.55\textwidth]{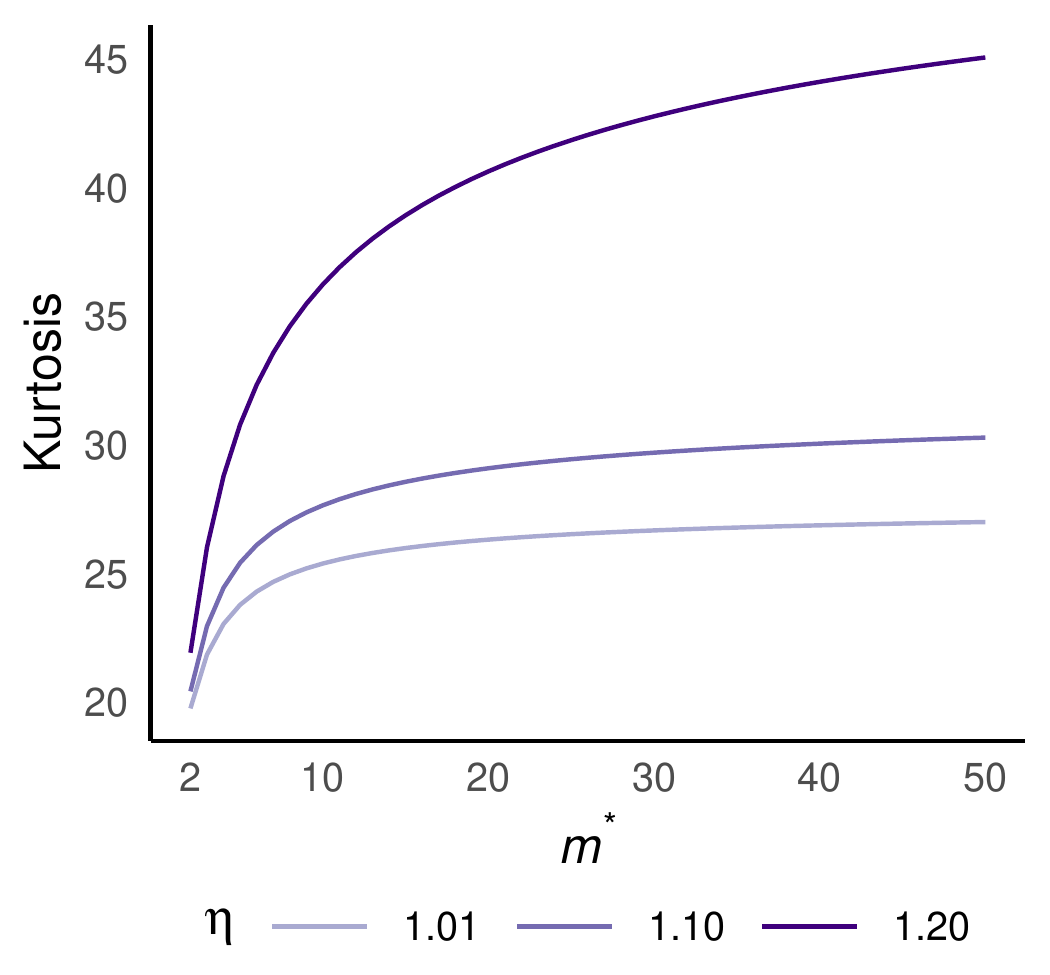}
        \caption{Pearson kurtosis of the cTDW distribution plotted against the shifted median $m^*$. The mixture weight was fixed at $\delta = 0.90$, while the tail-inflation parameter $\eta$ varied across the values shown in the legend. As $\eta$ increases, the contaminated component becomes more dispersed and heavier-tailed, resulting in higher kurtosis.}
        \label{fig:CTDW_kurtosis_eta}
    \end{figure}

\begin{figure}
        \centering
        \includegraphics[width=0.55\textwidth]{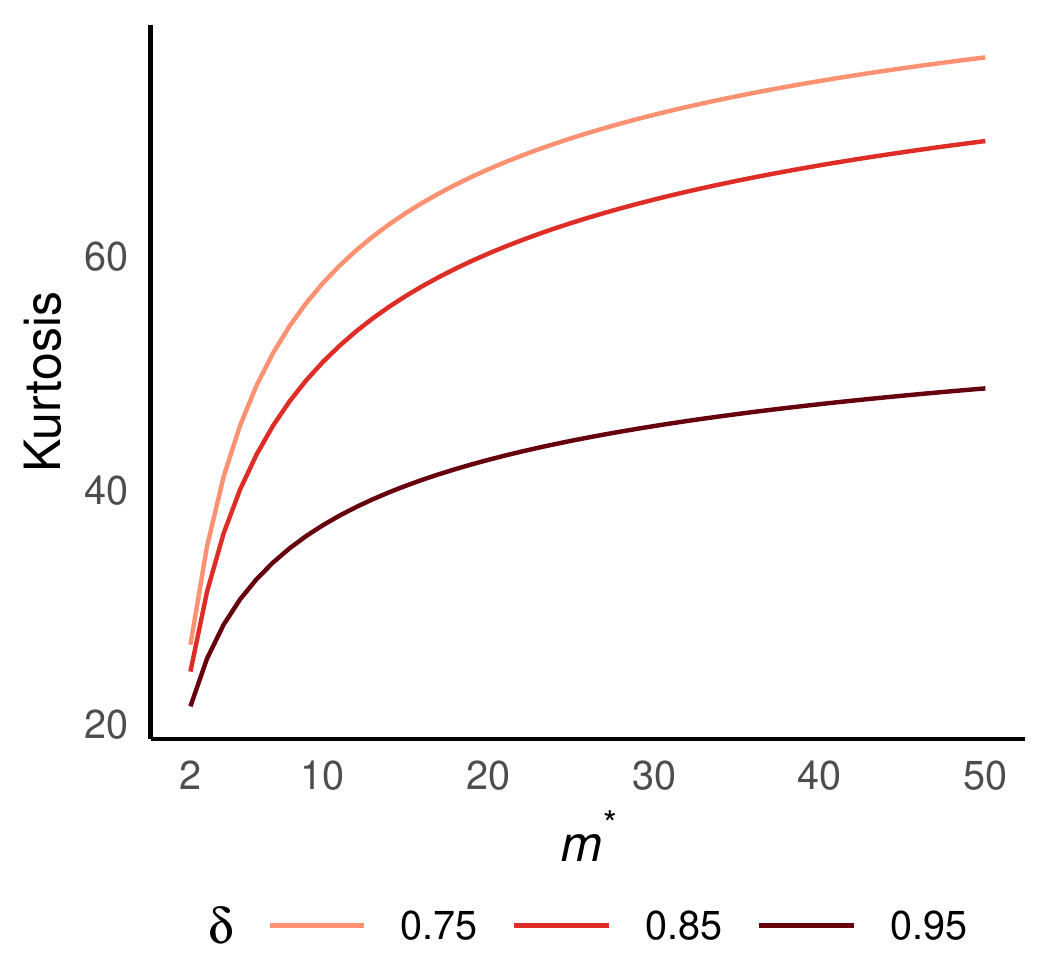}
        \caption{Pearson kurtosis of the cTDW distribution plotted against the shifted median $m^*$. The tail-inflation parameter was fixed at $\eta = 1.25$, while the mixture weight $\delta$ varied across the values shown in the legend. As $\delta$ increases, more weight is assigned to the narrower TDW component and less to the heavier-tailed component, resulting in lower kurtosis.}
        \label{fig:CTDW_kurtosis_delta}
    \end{figure}

    To illustrate the effect of contamination on tail behavior, \autoref{fig:CTDW_kurtosis_eta} and \autoref{fig:CTDW_kurtosis_delta} plot the Pearson kurtosis of the cTDW distribution against the shifted median $m^*$ for $c = 1$ and $\alpha = 1.5$. In \autoref{fig:CTDW_kurtosis_eta}, $\delta = 0.90$ is fixed and larger values of $\eta$ yield higher kurtosis. In \autoref{fig:CTDW_kurtosis_delta}, $\eta = 1.25$ is fixed and larger values of $\delta$ yield lower kurtosis. These plots confirm that the cTDW distribution provides interpretable control over tail heaviness and can accommodate an excess of large values relative to the single-component TDW model.

    Under this parameterization, $\delta$ represents the weight of the narrower component, whereas $1 - \delta$ is the weight of the heavier-tail component. In this paper, the main analysis follows the classical contamination convention, in which the heavier-tailed component is the minority component, whereas the alternative prior considered in the sensitivity analysis corresponds to a more general heavy-tailed two-component mixture interpretation. Throughout this paper, we therefore do not interpret $\delta$ itself as the contamination proportion or as an outlier proportion; rather, the term ``contaminated'' refers to the addition of a heavier-tailed component to the baseline TDW while preserving a common shifted median.
    
    The restriction $\eta > 1$ orders the two TDW components by tail thickness, with $P_{\mathrm{TDW}}\left(y \left|m^{*}, \eta \alpha, c\right.\right)$ having a more slowly decaying right tail than $P_{\mathrm{TDW}}\left(y \left|m^{*}, \alpha, c\right.\right)$. This ordering prevents label switching and improves practical identifiability by assigning a unique role to each component. However, it does not guarantee strong finite-sample identifiability, and estimation of $\eta$ may still be weak in mild-mixture settings. Throughout the paper, we therefore constrain the prior and the sampler to $\eta > 1$, which prevents label switching and ensures a unique ordering of the dispersion parameters.

    \subsection{Truncated negative binomial distribution}
    
    For context, we also consider NB models with lower truncation at $c$ \citep{SAMPFORD1955}. Let $X \sim \mathrm{NB}\left(\mu, \alpha\right)$ denote the NB distribution parameterized by mean $\mu>0$ and dispersion $\alpha>0$ (so $\operatorname{Var}\left(X\right)=\mu+\mu^{2}/\alpha$). Let $P_{\mathrm{NB}}\left(x \left|\mu, \alpha\right.\right)$ and $F_{\mathrm{NB}}\left(x \left|\mu, \alpha\right.\right)$ denote its PMF and CDF, with $F_{\mathrm{NB}}\left(x \left|\mu, \alpha\right.\right)=\sum_{k=0}^{x} P_{\mathrm{NB}}\left(k \left|\mu, \alpha\right.\right)$ for integers $x \ge 0$.
    
    Truncating at $c \in \left\{1,2,\ldots\right\}$ means defining
    \begin{equation}
        Y = X \left|\left(X \ge c\right)\right. .
    \end{equation}
    Denote the TNB PMF and CDF by $P_{\mathrm{TNB}}\left(y \left|\mu, \alpha, c\right.\right)$ and $F_{\mathrm{TNB}}\left(y \left|\mu, \alpha, c\right.\right)$; closed forms appear in {\color{red} Web} \autoref{sec:TNB_DIST} (Supporting Information). The truncated mean is
    \begin{equation}
        E\left(Y \left|\mu, \alpha, c\right.\right)
        = \frac{ \mu - \displaystyle\sum_{y=0}^{c-1} y P_{\mathrm{NB}}\left(y \left|\mu, \alpha\right.\right) }
               { 1 - F_{\mathrm{NB}}\left(c - 1 \left|\mu, \alpha\right.\right) }.
    \end{equation}
    Therefore, $E\left(Y \left|\mu, \alpha, c\right.\right)$ depends jointly on $\mu$ and the lower-tail mass through $F_{\mathrm{NB}}\left(c - 1 \left|\mu, \alpha\right.\right)$. Even with $\alpha$ and $c$ fixed, the mapping $\mu \mapsto E\left(Y \left|\mu, \alpha, c\right.\right)$ is nonlinear and does not, in general, admit a simple inverse. Thus, unlike the shifted-median parameterization used for the TDW/cTDW family, one cannot generally specify a target truncated mean and obtain $\mu$ through a convenient one-parameter regression link. In most settings, such calibration would require numerical inversion. Because this limitation is intrinsic to mean-centered NB models under truncation, a cTNB extension would inherit it and is not pursued here.

    \section{Modeling framework for count data} \label{sec:regression_models}

    \subsection{Regression model specification}

    Let each observation unit $i$ have a covariate vector $\mathbf{x}_i$, and let $\bm{\beta}$ be a regression coefficient vector. We link these covariates to the smooth shifted-median parameter $m_i^*$ by writing
    \begin{equation}
      \log\left(m_i^* - c\right) = \mathbf{x}_i^\top \bm{\beta}, \quad m_i^* = c + \exp\left(\mathbf{x}_i^\top \bm{\beta}\right), \quad m_i^* > c.
    \end{equation}
    Because $m_i^* - c$ must be strictly positive, the term $\exp\left(\mathbf{x}_i^\top \bm{\beta}\right)$ can span $\left(0, \infty\right)$ for any real-valued $\mathbf{x}_i^\top \bm{\beta}$. This ensures that the median exceeds the truncation point $c$ neatly.
    
    In the TDW model, the dispersion parameter $\alpha$ may remain fixed. Each observation $i$ thus has a TDW PMF,
    \begin{equation}
      P_{\mathrm{TDW}}\left(Y_i = y \left|m_i^*, \alpha, c\right.\right), \quad y \in \left\{c, c+1, \dots \right\}.
    \end{equation}
    When heavier tails are suspected, the cTDW model blends two TDW models that share the same median $m_i^*$ but differ in dispersion $\alpha$ vs. $\eta \alpha$ ($\eta>1$). Under this parameterization, $\delta$ denotes the mixture weight of the narrower component, whereas $1 - \delta$ is the weight of the heavier-tail component:
    \begin{equation}
      P_{\mathrm{cTDW}}\left(Y_i = y \left|m_i^*, \alpha, \eta, \delta, c\right.\right) = \delta P_{\mathrm{TDW}}\left(y \left|m_i^*, \alpha, c\right.\right) + \left(1-\delta\right) P_{\mathrm{TDW}}\left(y \left|m_i^*, \eta \alpha, c\right.\right).
    \end{equation}
    Although $\alpha$, $\delta$, or $\eta$ can also depend on covariates (via additional link functions), in this paper, we focus on modeling only $m_i^*$.
    
    Further details on model fitting and inference are provided in the next section.

    \subsection{Bayesian inference}

    In a Bayesian framework, all unknown parameters are treated as random variables. Let
    \begin{equation}
      \bm{\theta} =
      \begin{cases}
        \left(\bm{\beta}, \alpha\right), & \text{(TDW)}, \\
        \left(\bm{\beta}, \alpha, \eta, \delta\right), & \text{(cTDW)},
      \end{cases}
    \end{equation}
    where $\bm{\beta}$ is the regression coefficient vector, $\alpha$ is the baseline dispersion parameter, $\eta > 1$ scales the heavier-tail component, and $\delta$ is the mixture weight of the narrower component in the cTDW model. Given observed data $\mathcal{D}$, the posterior distribution of $\bm{\theta}$ is defined by
    \begin{equation}
      p\left(\bm{\theta} \left|\mathcal{D}\right.\right) \propto L\left(\mathcal{D} \left|\bm{\theta}\right.\right) p\left(\bm{\theta}\right),
    \end{equation}
    where $L\left(\mathcal{D} \left|\bm{\theta}\right.\right)$ is the likelihood for either the TDW or the cTDW model.
    
    We place independent priors on each component of $\bm{\theta}$. Specifically,
    \begin{equation}
      \bm{\beta} \sim \mathcal{N}\left(\mathbf{0}, \sigma_{\beta}^2 \mathbf{I}\right), \quad
      \alpha \sim \mathrm{Gamma}\left(a_{\alpha}, b_{\alpha}\right),
    \end{equation}
    \begin{equation}
      \eta \sim \mathrm{Gamma}\left(a_{\eta}, b_{\eta}\right) \mathbb{I}_{\left(1,\infty\right)}, \quad
      \delta \sim \mathrm{Uniform}\left(0.5, 1\right),
    \end{equation}
    where $\sigma_{\beta}^2$, $\left(a_{\alpha}, b_{\alpha}\right)$, and $\left(a_{\eta}, b_{\eta}\right)$ are hyperparameters, and $\mathbb{I}_{\left(1,\infty\right)}$ denotes truncation to $\left(1,\infty\right)$. Throughout, gamma priors are parameterized in terms of shape and rate. One may choose diffuse priors so that the data primarily drive parameter estimates, or more concentrated priors if strong domain knowledge is available. In the primary analysis, the prior $\delta \sim \mathrm{Uniform}\left(0.5, 1\right)$ reflects the convention that the narrower component should receive at least as much weight as the heavier-tail component. This is consistent with the standard interpretation of contamination in robust statistics, where the bulk component represents the majority of observations. To assess robustness to prior specification, we also fitted the cTDW model under the alternative prior combination
    \begin{equation}
      \eta \sim \mathrm{Uniform}\left(1, 10\right), \quad
      \delta \sim \mathrm{Uniform}\left(0, 1\right).
    \end{equation}
    This sensitivity analysis was motivated by the need to assess the influence of the prior specification on the contamination structure. In particular, the alternative prior on $\delta$ relaxes the majority-component restriction imposed in the primary analysis, while the alternative prior on $\eta$ avoids concentrating prior mass near the boundary $\eta = 1$, which would favor little or no tail inflation a priori. Thus, the sensitivity analysis examines whether posterior inference remains stable under a less restrictive specification for both the mixture weight and the tail-inflation parameter. Because the ordering constraint $\eta > 1$ still distinguishes the components by tail thickness, the alternative prior on $\delta$ affects interpretation rather than component labeling.
    
    To carry out Bayesian inference, we encode the TDW or cTDW likelihood in JAGS together with the specified priors. Using Markov chain Monte Carlo (MCMC) sampling, JAGS generates samples from the posterior $p\left(\bm{\theta} \left|\mathcal{D}\right.\right)$. Once convergence is verified (e.g., via trace plots or the Brooks-Gelman-Rubin potential scale reduction factor (PSRF) \citep{gelman1992inference, BROOKS1998}), we compute summary statistics such as posterior medians and Bayesian credible intervals (BCIs) directly from the MCMC samples.
    
    \section{Model adequacy checks} \label{sec:model_adequacy}

    We evaluate model adequacy using the \textsf{DHARMa} package \citep{hartig2024package}, which generates simulation-based quantile residuals from the posterior predictive distribution. Specifically, parameter draws are taken from the fitted posterior, and each draw is used to simulate new responses for every observation. Under correct model specification, these residuals should approximate a $\mathrm{Uniform}\left(0,1\right)$ distribution. In a uniform quantile-quantile (QQ) plot, the 1:1 diagonal (i.e., the line $y=x$) indicates perfect agreement between the empirical and theoretical quantiles; thus, residuals lying close to this line suggest a well-fitting model.

    In addition to the residual checks, we perform a K-L divergence assessment to detect influential observations. Following Wang \& Luo \citep{WANG2016}, we approximate how much the posterior distribution changes when each data point is omitted. In the cTDW model, posterior probabilities of assignment to the heavier-tail component provide a natural complementary diagnostic at the observation level. However, these probabilities quantify component allocation rather than posterior influence and therefore answer a different question from the K-L divergence. In particular, an observation may have a high posterior probability of belonging to the heavier-tail component but still have limited influence on inference because the cTDW is designed to accommodate such observations. We therefore use the K-L divergence as our primary influence diagnostic, since it can also be applied uniformly to the TDW and TNB models. \ref{sec:KL_METHOD} provides the derivations and the threshold we use for classifying potentially influential observations.

    \section{Model comparison via leave-one-out cross-validation} \label{sec:model_comparison}

    To compare the predictive performance of different models, we employ LOO cross-validation through the \textsf{loo} package \citep{vehtari2024package}. This approach approximates out-of-sample predictive accuracy by using pointwise log-likelihood contributions for each observation, evaluated across posterior draws from the full-data fit, together with Pareto-smoothed importance sampling (PSIS) to approximate the LOO distribution. Specifically, we extract the log-likelihood for each posterior draw and observation from the full-data fit, then apply PSIS to stabilize estimates of the LOO predictive densities. The resulting LOOIC and associated metrics (e.g., the effective number of parameters) are used to identify which model provides a better predictive fit. Lower LOOIC values indicate superior generalization to new data.

    \section{Application to hospital length-of-stay data}
    \label{sec:application}
    
    \subsection{Implementation and model specification}
    
    We apply both the single-component TDW and the cTDW models described in \autoref{sec:regression_models} to the Arizona hospital LOS dataset \citep{hilbe2011negative, RCOUNT2025}. Each observation $i$ has LOS $Y_i \in \left\{1,2,\dots \right\}$ and covariates: procedure type (CABG or PTCA), admission category (elective or urgent/emergent), and sex (female or male). We encode these as dummy variables in the design matrix $\mathbf{x}_i$ so that our regression vector $\bm{\beta}$ has the following interpretation:
    \begin{equation}
        \begin{aligned}
          \beta_0 &= \text{Intercept}, \\
          \beta_1 &= \text{CABG vs. PTCA indicator}, \\
          \beta_2 &= \text{Urgent/emergent vs. elective indicator}, \\
          \beta_3 &= \text{Male vs. female indicator}, \\
          \beta_4 &= \text{CABG $\times$ urgent/emergent interaction}, \\
          \beta_5 &= \text{CABG $\times$ male interaction}, \\
          \beta_6 &= \text{Urgent/emergent $\times$ male interaction}, \\
          \beta_7 &= \text{CABG $\times$ urgent/emergent $\times$ male (three-way interaction)}.
        \end{aligned}
    \end{equation}
    In both the TDW and cTDW fits, we use the shifted-median link:
    \begin{equation}
      \log\left(m_i^* - 1\right) = \mathbf{x}_i^\top \bm{\beta}, \quad m_i^* > 1.
    \end{equation}
    For the TDW model, each observation $i$ follows
    \begin{equation}
        P_{\mathrm{TDW}}\left(Y_i = y \left|m_i^*, \alpha, c = 1\right.\right),
    \end{equation}
    and for the cTDW model,
    \begin{equation}
        \begin{aligned}
          P_{\mathrm{cTDW}}\left(Y_i = y \left|m_i^*, \alpha, \eta, \delta, c = 1\right.\right) 
          &= \delta P_{\mathrm{TDW}}\left(y \left|m_i^*, \alpha, c = 1\right.\right) \\
          &\quad + \left(1-\delta\right) P_{\mathrm{TDW}}\left(y \left|m_i^*, \eta \alpha, c = 1\right.\right),
        \end{aligned}
    \end{equation}
    as described in \autoref{sec:regression_models}. Here, $\alpha$ is the baseline dispersion parameter, $\eta > 1$ scales the heavier-tail component, and $\delta$ is the mixture weight assigned to the narrower component, so that $1 - \delta$ is the weight of the heavier-tail component. We set $c=1$ to reflect the minimum possible hospital stay for this dataset, which is one day.
    
    We assign priors as follows:
    \begin{equation}
      \beta_j \sim \mathcal{N}\left(0, 10^3\right), \quad \alpha \sim \mathrm{Gamma}\left(0.001, 0.001\right),
    \end{equation}
    \begin{equation}
      \eta \sim \mathrm{Gamma}\left(0.001, 0.001\right) \mathbb{I}_{\left(1,\infty\right)}, \quad \delta \sim \mathrm{Uniform}\left(0.5, 1\right),
    \end{equation}
    where $j = 0,1,\ldots,7$, and the gamma distributions are parameterized in terms of shape and rate, consistent with the JAGS implementation. The prior restriction on $\delta$ reflects the modeling convention used in the primary analysis that the narrower component receives at least as much total weight as the heavier-tail component.
    
    All models are run in JAGS (via the \texttt{runjags} package \citep{DENWOOD2016}) using four chains, each with 2000 adaptation steps and 4000 burn-in steps. After burn-in, we run each chain for 25~000 iterations, thinning every 5 draws, yielding 5000 post-burn-in samples per chain. Combining the four chains results in 20~000 total posterior draws for each parameter. Convergence is assessed via the PSRF, which remains at or below 1.05 for all parameters.

    We assessed model fit using simulation-based residual diagnostics from the \texttt{DHARMa} package \citep{hartig2024package} based on 500 posterior predictive draws (see \autoref{sec:model_adequacy}). We compared models using LOO cross-validation in the \texttt{loo} package \citep{vehtari2024package} (see \autoref{sec:model_comparison}) and identified influential observations by computing the K-L divergence from MCMC-based log-likelihood estimates (also described in \autoref{sec:model_adequacy}).

    We fit a TNB model using the same linear predictor and log link as in the DW models, applied to the untruncated NB mean,
    \begin{equation}
        \log\left(\mu_i\right) = \mathbf{x}_i^\top \bm{\beta}.
    \end{equation}
    The dispersion parameter used the same gamma prior as the TDW dispersion. Because the link targets the mean of the untruncated distribution while the data are truncated at $c$, there is no simple way to interpret coefficients for the truncated outcome; see {\color{red} Web} \autoref{sec:TNB_DIST}. We therefore report the TNB solely as a predictive benchmark. Posterior sampling and diagnostics followed the same workflow as for the TDW/cTDW fits.

    All R code for replicating the results is available on GitHub at \href{https://github.com/DABURGER1/Robust-Count-cTDW}{Robust-Count-cTDW}.

    \subsection{Residual checks and leave-one-out cross-validation}

\begin{figure}
        \centering
        \subfloat[TDW]{
            \includegraphics[width=0.7\textwidth]{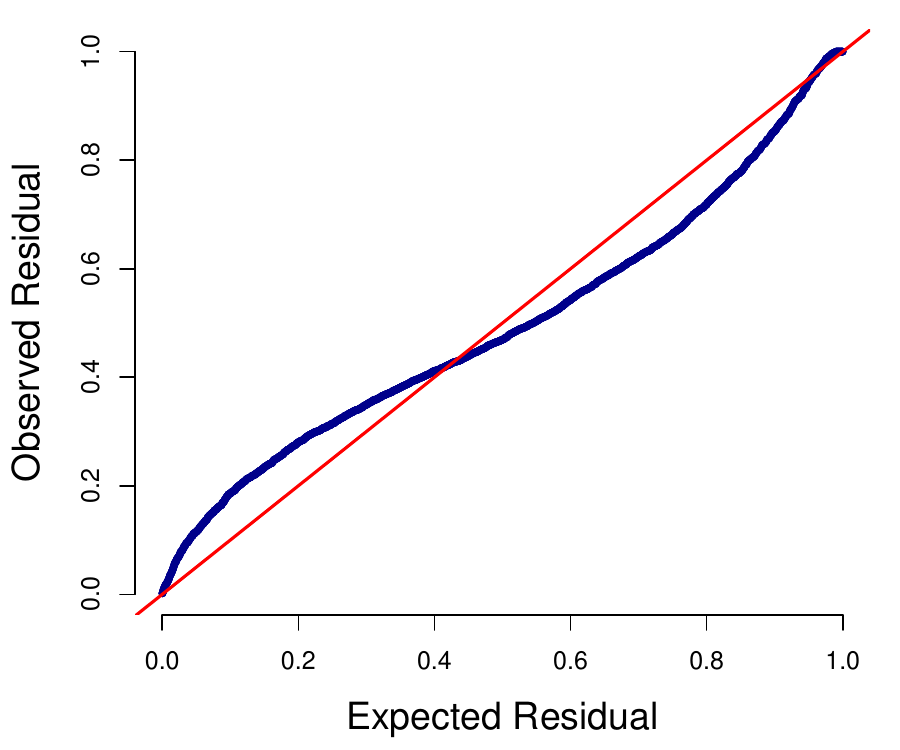}
        }
        \hfill
        \subfloat[cTDW]{
            \includegraphics[width=0.7\textwidth]{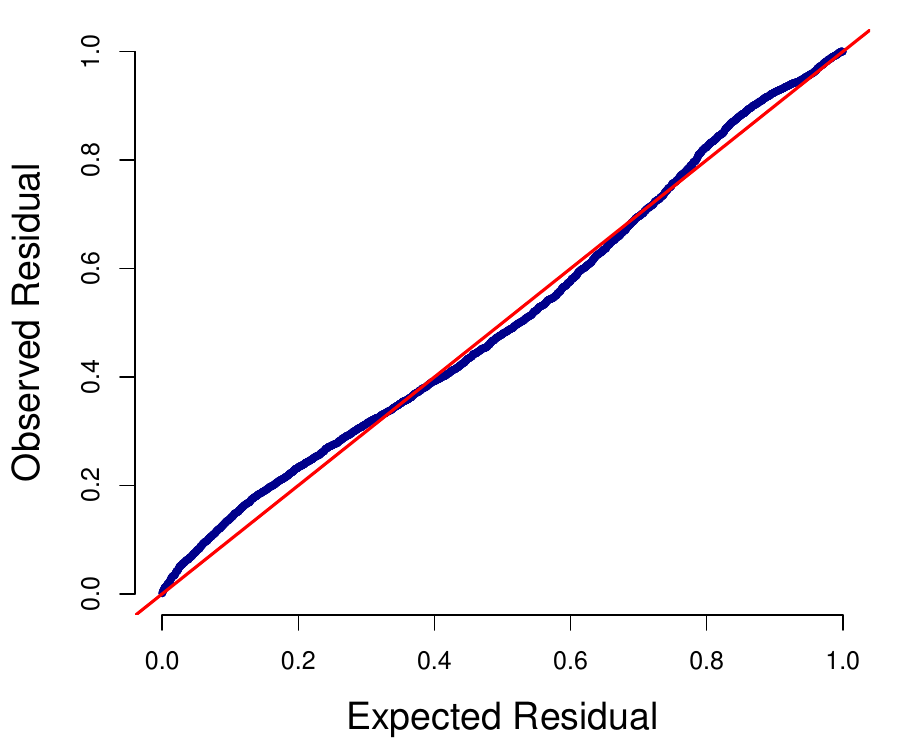}
        }
        \caption{QQ plots of simulation-based residuals for (a) the single-component TDW model and (b) the cTDW model. The blue points show the empirical distribution of residuals against the ideal uniform distribution (red diagonal). Compared to the single-component TDW model, the cTDW residuals adhere more closely to the diagonal, indicating improved model fit.}
        \label{fig:LOS_residuals}
    \end{figure}

    \autoref{fig:LOS_residuals} shows the simulation-based residual diagnostics for the TDW and cTDW models. The cTDW residuals align more closely with the uniform diagonal than those from the TDW model. LOO cross-validation likewise favors the cTDW, with $\mathrm{LOOIC}\approx19~541$, compared with $\mathrm{LOOIC}\approx19~750$ for the TNB, and $\mathrm{LOOIC}\approx20~304$ for the TDW. As the model with the smallest LOOIC is considered the preferred one, the cTDW yields the best expected predictive performance. Residual checks for the TNB show departures similar to the TDW (see Supporting Information, {\color{red} Web} \autoref{sec:WEB_FIGURES}; QQ plot in {\color{red} Web} \autoref{fig:LOS_TNB_residuals}).

    \subsection{Influential observations}

\begin{figure}
        \centering
        \subfloat[TDW]{
            \includegraphics[width=0.75\textwidth]{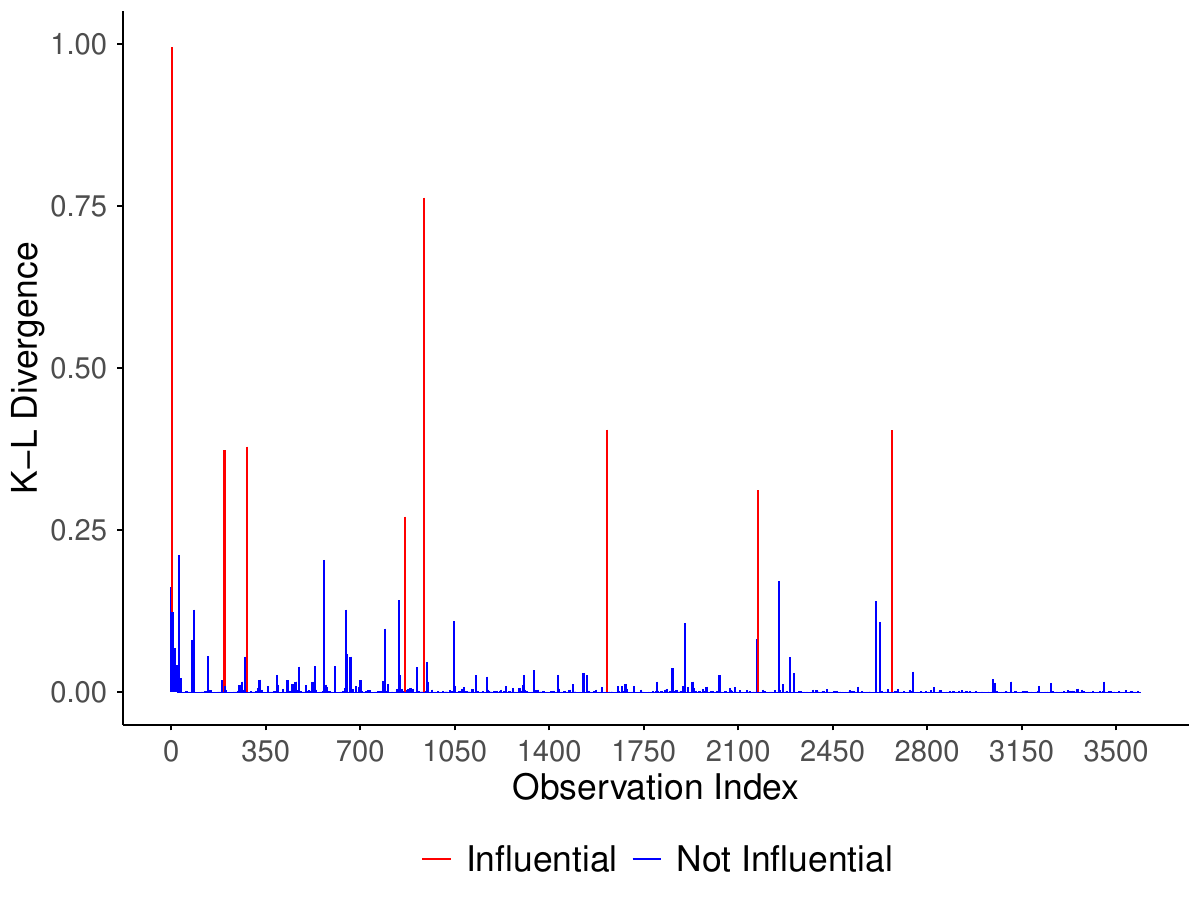}
        }
        \hfill
        \subfloat[cTDW]{
            \includegraphics[width=0.75\textwidth]{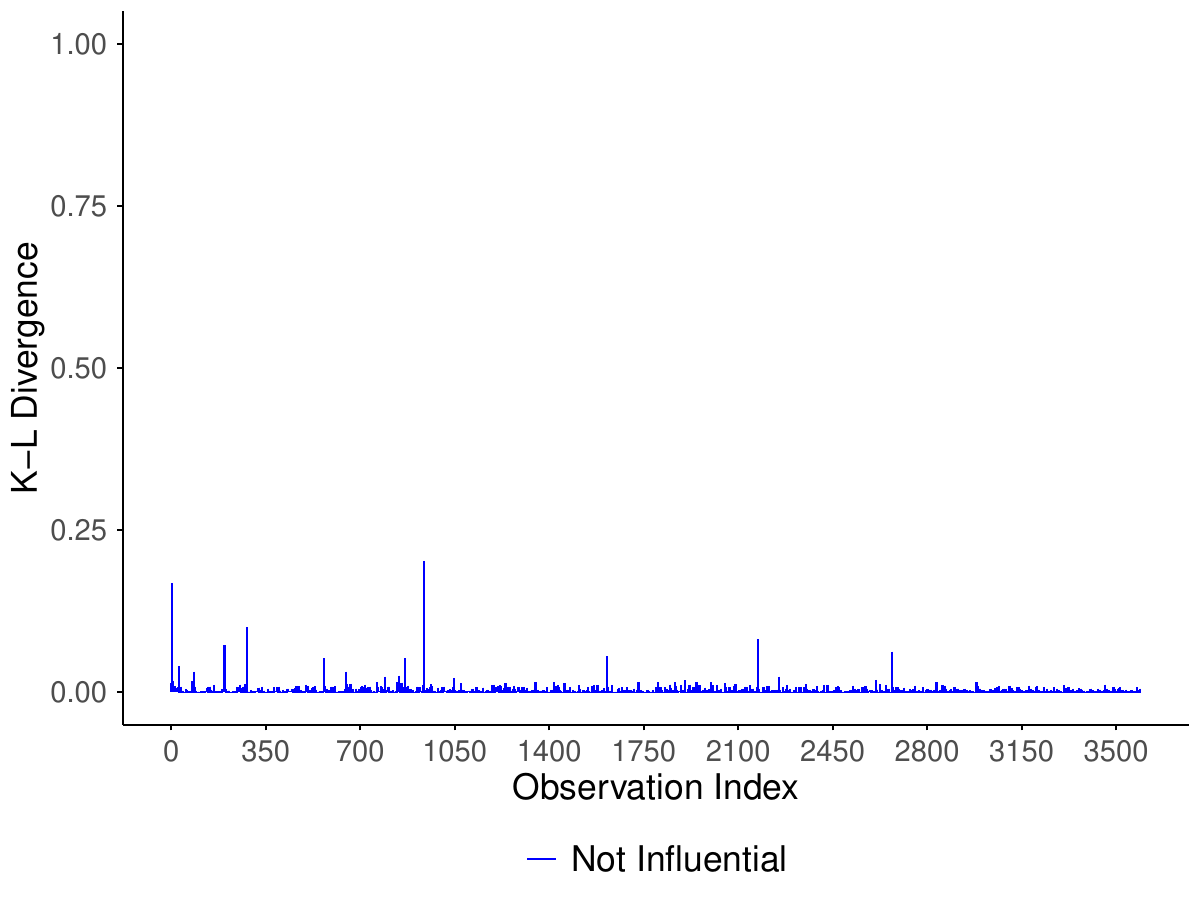}
        }
        \caption{K-L divergence plots for (a) the TDW and (b) the cTDW models. Each vertical bar represents an observation, colored red if classified as influential and blue otherwise. Under the single-component TDW model, several points exhibit disproportionately large influence, whereas the cTDW mixture reduces these outliers by better accommodating heavy-tailed behavior.}
        \label{fig:LOS_KL}
    \end{figure}

    \autoref{fig:LOS_KL} presents the K-L divergence plots for the TDW and cTDW models. Fewer observations exceed the K-L threshold for being considered influential under the cTDW model, indicating that the mixture framework better accommodates high-influence data points than the single-component TDW model.

    For completeness, {\color{red} Web} \autoref{fig:LOS_TNB_KL} (Supporting Information, {\color{red} Web} \autoref{sec:WEB_FIGURES}) shows that the TNB flags comparatively few influential observations; however, its residual departures resemble those of the TDW, and its predictive fit is intermediate according to LOOIC.

    \subsection{Posterior estimates and median length-of-stay}

    \autoref{fig:median_bands_DW_cDW} shows the fitted shifted medians $m_i^*$ for each subgroup, comparing the single-component TDW and the cTDW models. Overall, the TDW tends to estimate slightly higher medians in some groups, presumably because extreme stays inflate the single-component fit. By contrast, the cTDW accommodates the heavy upper tail through its mixture component, yielding a slightly lower or tighter center in most subgroups.
    
    \autoref{tab:median_bands_DW_cDW} reports posterior medians and 95\% BCIs for $\left\{\beta_j\right\}$, $\alpha$, $\eta$, and $\delta$. The single-component TDW yields $\alpha \approx 0.58$, whereas the cTDW estimates $\alpha \approx 0.30$ for the narrower component, $\eta \approx 2.85$ for the heavier-tail inflation factor, and $\delta \approx 0.69$ for the mixture weight of the narrower component. Thus, most of the total mass is assigned to the narrower component, while a smaller, heavier-tail component accommodates extreme counts. To assess sensitivity to the prior specification, we also fitted the alternative cTDW model with $\delta \sim \mathrm{Uniform}\left(0,1\right)$ and $\eta \sim \mathrm{Uniform}\left(1,10\right)$. Results were very similar under this alternative specification, with broadly unchanged posterior summaries, subgroup-specific shifted medians, and 95\% BCIs. This suggests that the fitted medians and substantive conclusions are not materially driven by the prior assumptions in the main analysis. Full sensitivity results appear in the Supporting Information: {\color{red} Web} \autoref{fig:median_bands_cDW} ({\color{red} Web} \autoref{sec:WEB_FIGURES}) displays subgroup medians and 95\% BCIs under both prior specifications, and {\color{red} Web} \autoref{tab:median_bands_cDW} ({\color{red} Web} \autoref{sec:WEB_TABLES}) reports the corresponding parameter summaries.

    \begin{longtable}{
        l 
        S[table-format=3.3] 
        S[table-format=3.3] 
        S[table-format=2.3] 
        l 
        S[table-format=3.3] 
        S[table-format=3.3] 
        S[table-format=2.3]}
        \caption{Posterior medians (Est.) and 95\% BCIs for the single-component TDW and cTDW models applied to the Arizona hospital LOS data. The regression coefficients $\beta_j$ link covariates to a shifted median via $\log\left(m_i^* - 1\right) = \mathbf{x}_i^\top \bm{\beta}$, where $\beta_1$ through $\beta_7$ capture the effects of procedure type (CABG vs. PTCA), admission category (urgent/emergent vs. elective), and sex (male vs. female), including interactions. The dispersion parameter is $\alpha$ in both models, whereas $\eta$ and $\delta$ appear only in the cTDW model, with $\eta$ inflating the heavier-tail subdistribution and $\delta$ denoting the mixture weight of the narrower component.} \label{tab:median_bands_DW_cDW} \\
            
        \hline
        & \multicolumn{7}{c}{\textbf{Model}} \\
        \cline{2-8}
        & \multicolumn{3}{c}{\textbf{TDW}} & & \multicolumn{3}{c}{\textbf{cTDW}} \\
        \cline{2-4} \cline{6-8}
        \textbf{Parameter} & \textbf{Est.} & \multicolumn{2}{c}{\textbf{95\% BCI}} & & \textbf{Est.} & \multicolumn{2}{c}{\textbf{95\% BCI}} \\
        \hline
        \endfirsthead
            
        \multicolumn{8}{c}{{\bfseries Table \thetable\ (Continued)}}\\
        \hline
        & \multicolumn{7}{c}{\textbf{Model}} \\
        \cline{2-8}
        & \multicolumn{3}{c}{\textbf{TDW}} & & \multicolumn{3}{c}{\textbf{cTDW}} \\
        \cline{2-4} \cline{6-8}
        \textbf{Parameter} & \textbf{Est.} & \multicolumn{2}{c}{\textbf{95\% BCI}} & & \textbf{Est.} & \multicolumn{2}{c}{\textbf{95\% BCI}} \\
        \hline
        \endhead
            
        \hline
        \multicolumn{8}{r}{{Continued on next page}}\\
        \endfoot
            
        \hline
        \endlastfoot
            
          $\beta_{0}$ & 1.023 & [0.926; & 1.117] &  & 0.851 & [0.765; & 0.942] \\* 
          $\beta_{1}$ & 1.283 & [1.152; & 1.418] &  & 1.444 & [1.333; & 1.553] \\* 
          $\beta_{2}$ & 0.722 & [0.613; & 0.836] &  & 0.871 & [0.767; & 0.974] \\* 
          $\beta_{3}$ & -0.106 & [-0.221; & 0.015] &  & -0.090 & [-0.198; & 0.015] \\* 
          $\beta_{4}$ & -0.457 & [-0.619; & -0.300] &  & -0.604 & [-0.734; & -0.470] \\* 
          $\beta_{5}$ & -0.005 & [-0.166; & 0.151] &  & -0.016 & [-0.144; & 0.116] \\* 
          $\beta_{6}$ & -0.072 & [-0.213; & 0.066] &  & -0.047 & [-0.176; & 0.080] \\* 
          $\beta_{7}$ & 0.046 & [-0.146; & 0.243] &  & 0.020 & [-0.140; & 0.180] \\ 
          $\alpha$ & 0.584 & [0.569; & 0.599] &  & 0.303 & [0.280; & 0.326] \\* 
          $\eta$ & & & &  & 2.846 & [2.655; & 3.053] \\* 
          $\delta$ & & & &  & 0.688 & [0.631; & 0.741] \\ 
        \hline
    \end{longtable}
    
    \afterpage{
        \begin{landscape}
        \pagestyle{landscape}
            \begin{figure}[htbp]
                \centering
                \includegraphics[scale = 0.8]{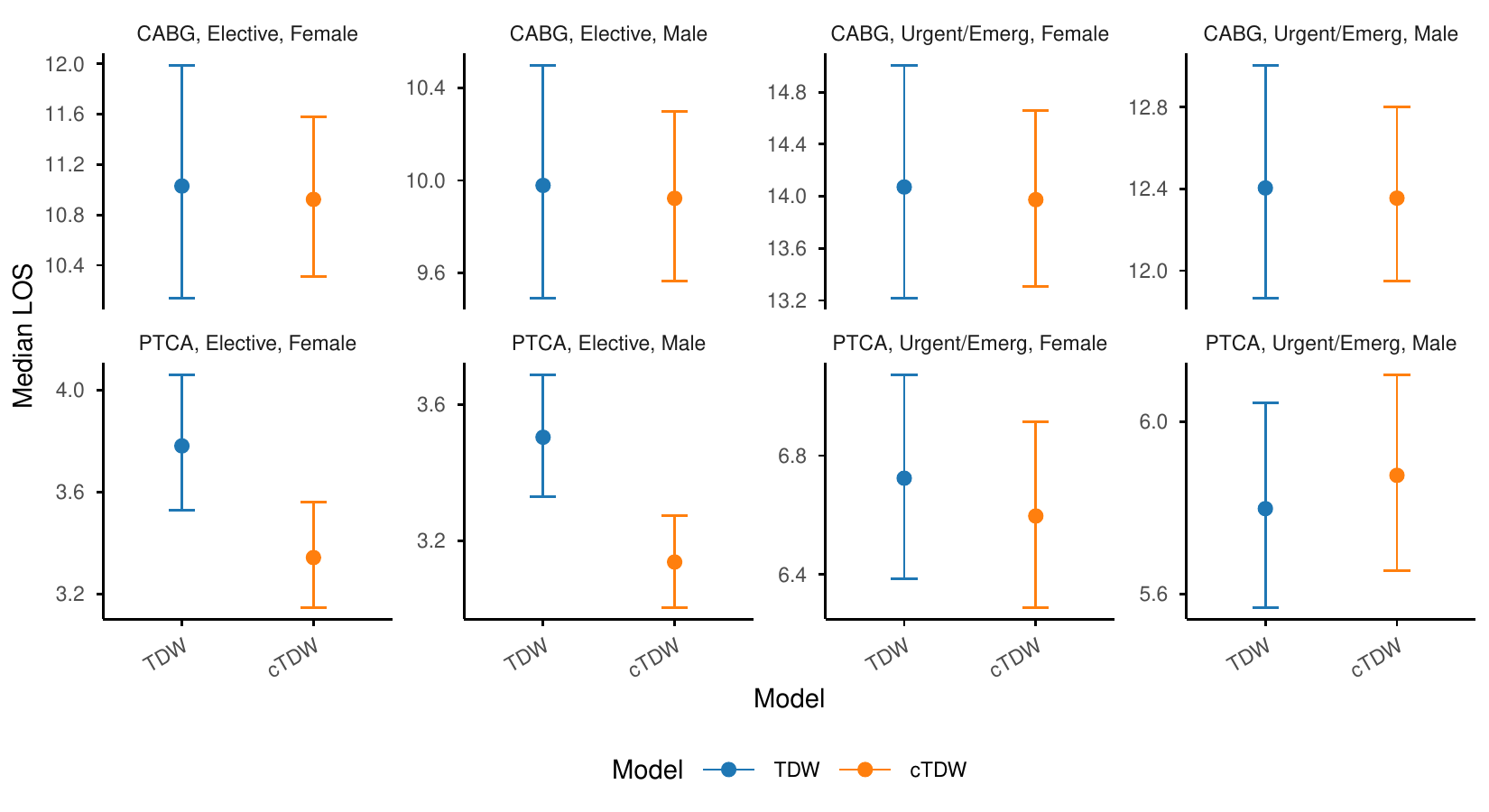}
                \caption{Posterior estimates of the shifted median $m^*$ for each subgroup, modeled under the single-component TDW and cTDW models. Each point represents the posterior median of $m^*$, and the vertical bars (whiskers) show the corresponding 95\% BCI. Facets distinguish the combinations of procedure type, admission category, and patient sex. The TDW estimates (blue) and cTDW estimates (orange) are displayed side by side for each subgroup.}
                \label{fig:median_bands_DW_cDW}
            \end{figure}
        \end{landscape}
    }

    \section{Simulation study} \label{sec:simulation}

    The simulation study serves two purposes. First, it evaluates the operating characteristics of the proposed cTDW model when the data are generated from the cTDW family itself. Second, it assesses the robustness advantage of the cTDW relative to the single-component TDW when the true data-generating mechanism is a genuinely heavy-tailed two-component mixture. Thus, the study is not limited to parameter recovery under the proposed model, but also examines the effect of fitting the simpler TDW model when the data arise from mixture-generated heavy-tailed settings.

    To investigate the performance of the proposed cTDW model and compare it with the single-component TDW model, we conducted a simulation study under several cTDW mixture settings. We considered the single-covariate regression model
    \begin{equation}
        \log\left(m_i^* - c\right) = \beta_0 + \beta_1 x_i,
    \end{equation}
    where $m_i^*$ denotes the shifted median for observation $i$, $c$ is the lower bound, and $\beta_0$ and $\beta_1$ are regression coefficients. In each replicate, the covariates were generated independently as
    \begin{equation}
        X_i \overset{\mathrm{iid}}{\sim} \mathcal{N}\left(0, 1\right), \quad i = 1,\ldots,N,
    \end{equation}
    and response counts $Y_i$ were then generated from the cTDW model with parameters $\left\{\beta_0, \beta_1, \alpha, \eta, \delta\right\}$, where $\delta$ denotes the weight of the narrower component and $1 - \delta$ the weight of the heavier-tail component. We then fitted both the single-component TDW and cTDW models to each generated dataset to assess the effect of fitting a single-component model when the true data-generating mechanism is a mixture.

    For each lower bound setting $c \in \left\{0, 1\right\}$, for each sample size $N \in \left\{25, 50, 100, 200\right\}$, and for each mixture setting $\left(\eta, \delta\right) \in \left\{\left(2, 0.90\right), \left(5, 0.55\right), \left(10, 0.55\right)\right\}$, we generated 500 independent datasets. Here and throughout, $\delta$ denotes the weight of the narrower component, so that the corresponding heavier-tail weights are $1 - \delta \in \left\{0.10, 0.45, 0.45\right\}$. We fixed the true values at $\beta_0 = 2$, $\beta_1 = 0.3$, and $\alpha = 0.6$. In total, this yielded 24 simulation scenarios, arising from four sample sizes, two truncation limits, and three mixture settings ($4 \times 2 \times 3 = 24$).

    We used the \verb|autorun.jags| feature of the \texttt{runjags} package \citep{DENWOOD2016} to ensure convergence across all simulation replicates.

    For the cTDW model, we summarized performance across the 500 replicates using mean bias and coverage probability (CP) of the 95\% BCIs for all parameters; these results are reported in \autoref{tab:sim_results_cTDW}. For the regression coefficients $\beta_0$ and $\beta_1$, we additionally report root mean squared error (RMSE) and average 95\% BCI length for both the TDW and cTDW models; these summaries are given in \autoref{tab:sim_result_beta}.

    \autoref{tab:sim_results_cTDW} reports the mean bias and CPs for all parameters under the cTDW model. Overall, the regression coefficients $\beta_0$ and $\beta_1$ are recovered well across most scenarios, with biases generally close to zero and CPs often near the nominal level. By contrast, the mixture-tail parameters are more difficult to estimate, especially in smaller samples and under truncation at $c = 1$. In particular, $\eta$ can exhibit substantial positive bias in some scenarios, indicating that finite-sample identification of the heavier-tail component may be weak even under the ordering constraint $\eta > 1$. A plausible explanation for the poorer behavior under $c = 1$ is the loss of lower-tail information through truncation. As shown in {\color{red} Web} \autoref{fig:ComponentSeparation} (Supporting Information), the zero cell provides information that helps distinguish the narrower and heavier-tail components, and this information is no longer available once the distribution is truncated at $c = 1$. Consequently, finite-sample identification of $\alpha$, $\eta$, and $\delta$ may become weaker. The dispersion parameter $\alpha$ also becomes harder to estimate in the more strongly heavy-tailed settings, particularly when $c = 1$.

    \autoref{tab:sim_result_beta} compares the simulation summaries for the regression coefficients $\beta_0$ and $\beta_1$ under the TDW and cTDW models. Under the mild mixture setting $\left(\eta, \delta\right) = \left(2, 0.90\right)$, the two models perform similarly, which is expected because the heavier-tail component has relatively little weight. However, under the more pronounced mixture settings $\left(\eta, \delta\right) = \left(5, 0.55\right)$ and $\left(10, 0.55\right)$, the single-component TDW model shows substantially larger RMSE values, wider BCIs, and poorer coverage than the cTDW model. These differences are especially marked for $\beta_1$ when $c = 0$. Similar, although somewhat less pronounced, improvements are observed under truncation at $c = 1$. Overall, the results show that the cTDW model is considerably more robust than the single-component TDW model when the data arise from a genuinely heavy-tailed mixture.

    \setlength{\tabcolsep}{7pt}

    \begin{longtable}{S[table-format=2.0]S[table-format=2.2]S[table-format=3.0]lS[table-format=1.2]S[table-format=2.4]S[table-format=1.3]lS[table-format=2.4]S[table-format=1.3]}
    \caption{Simulation results for the cTDW model across all scenarios. For each combination of $c$, $\eta$, $\delta$, and sample size $N$, 500 datasets were generated. Here, $\delta$ denotes the weight of the narrower component, so that the heavier-tail component has weight $1 - \delta$. The table reports the mean bias and 95\% BCI coverage probability (CP) for each parameter.}
    \label{tab:sim_results_cTDW} \\
    \hline
    & & & & & \multicolumn{2}{c}{$\bm{c=0}$} & & \multicolumn{2}{c}{$\bm{c=1}$} \\
    \cline{6-7} \cline{9-10}
    $\bm\eta$ & $\bm\delta$ & {$\bm N$} & {\bf Parameter} & {\bf Value} & {\bf Bias} & {\bf CP} & & {\bf Bias} & {\bf CP} \\ \hline
    \endfirsthead
    
    \multicolumn{10}{c}
    {{\bfseries \tablename\ \thetable{} -- continued from previous page}} \\
    \hline
    & & & & & \multicolumn{2}{c}{$\bm{c=0}$} & & \multicolumn{2}{c}{$\bm{c=1}$} \\
    \cline{6-7} \cline{9-10}
    $\bm\eta$ & $\bm\delta$ & {$\bm N$} & {\bf Parameter} & {\bf Value} & {\bf Bias} & {\bf CP} & & {\bf Bias} & {\bf CP} \\ \hline
    \endhead
    
    \hline \multicolumn{10}{r}{{Continued on next page}} \\
    \endfoot
    
    \hline
    \endlastfoot
    
      2 & 0.9 & 25 & $\beta_0$ & 2.00 & 0.0014 & 0.936 &  & -0.0075 & 0.948 \\* 
       &  &  & $\beta_1$ & 0.30 & 0.0036 & 0.950 &  & 0.0089 & 0.964 \\* 
       &  &  & $\alpha$ & 0.60 & -0.0466 & 0.960 &  & -0.0311 & 0.966 \\* 
       &  &  & $\eta$ & 2.00 & 1.5068 & 0.982 &  & 4.8236 & 0.978 \\* 
       &  &  & $\delta$ & 0.90 & -0.0721 & 0.990 &  & -0.0753 & 0.998 \\ 
      2 & 0.9 & 50 & $\beta_0$ & 2.00 & 0.0067 & 0.944 &  & -0.0028 & 0.944 \\* 
       &  &  & $\beta_1$ & 0.30 & -0.0008 & 0.946 &  & 0.0019 & 0.952 \\* 
       &  &  & $\alpha$ & 0.60 & -0.0402 & 0.958 &  & -0.0099 & 0.974 \\* 
       &  &  & $\eta$ & 2.00 & 0.4802 & 0.974 &  & 1.3380 & 0.972 \\* 
       &  &  & $\delta$ & 0.90 & -0.0775 & 0.990 &  & -0.0479 & 1.000 \\ 
      2 & 0.9 & 100 & $\beta_0$ & 2.00 & 0.0021 & 0.954 &  & -0.0070 & 0.950 \\* 
       &  &  & $\beta_1$ & 0.30 & 0.0032 & 0.954 &  & 0.0029 & 0.942 \\* 
       &  &  & $\alpha$ & 0.60 & -0.0204 & 0.966 &  & -0.0106 & 0.964 \\* 
       &  &  & $\eta$ & 2.00 & 0.0203 & 0.978 &  & 0.5848 & 0.974 \\* 
       &  &  & $\delta$ & 0.90 & -0.0806 & 0.990 &  & -0.0518 & 0.998 \\ 
      2 & 0.9 & 200 & $\beta_0$ & 2.00 & -0.0041 & 0.960 &  & 0.0013 & 0.934 \\* 
       &  &  & $\beta_1$ & 0.30 & -0.0011 & 0.952 &  & 0.0012 & 0.946 \\* 
       &  &  & $\alpha$ & 0.60 & -0.0187 & 0.958 &  & -0.0148 & 0.962 \\* 
       &  &  & $\eta$ & 2.00 & -0.1001 & 0.982 &  & 0.1410 & 0.978 \\* 
       &  &  & $\delta$ & 0.90 & -0.0859 & 0.966 &  & -0.0657 & 0.994 \\ 
      5 & 0.55 & 25 & $\beta_0$ & 2.00 & 0.0146 & 0.956 &  & -0.0314 & 0.966 \\* 
       &  &  & $\beta_1$ & 0.30 & 0.0008 & 0.936 &  & -0.0130 & 0.962 \\* 
       &  &  & $\alpha$ & 0.60 & 0.3315 & 0.920 &  & 0.6867 & 0.852 \\* 
       &  &  & $\eta$ & 5.00 & 0.2480 & 0.996 &  & 11.9290 & 0.972 \\* 
       &  &  & $\delta$ & 0.55 & 0.1571 & 0.968 &  & 0.1913 & 0.994 \\ 
      5 & 0.55 & 50 & $\beta_0$ & 2.00 & -0.0176 & 0.944 &  & -0.0343 & 0.934 \\* 
       &  &  & $\beta_1$ & 0.30 & -0.0071 & 0.950 &  & -0.0166 & 0.948 \\* 
       &  &  & $\alpha$ & 0.60 & 0.1536 & 0.894 &  & 0.2852 & 0.838 \\* 
       &  &  & $\eta$ & 5.00 & -0.0799 & 0.978 &  & 6.5577 & 0.964 \\* 
       &  &  & $\delta$ & 0.55 & 0.0918 & 0.950 &  & 0.1651 & 0.960 \\ 
      5 & 0.55 & 100 & $\beta_0$ & 2.00 & -0.0148 & 0.942 &  & -0.0138 & 0.952 \\* 
       &  &  & $\beta_1$ & 0.30 & -0.0066 & 0.934 &  & -0.0101 & 0.934 \\* 
       &  &  & $\alpha$ & 0.60 & 0.0498 & 0.944 &  & 0.1436 & 0.870 \\* 
       &  &  & $\eta$ & 5.00 & 0.0145 & 0.964 &  & 1.8726 & 0.972 \\* 
       &  &  & $\delta$ & 0.55 & 0.0463 & 0.952 &  & 0.1036 & 0.962 \\ 
      5 & 0.55 & 200 & $\beta_0$ & 2.00 & -0.0104 & 0.940 &  & -0.0039 & 0.948 \\* 
       &  &  & $\beta_1$ & 0.30 & -0.0010 & 0.958 &  & 0.0000 & 0.938 \\* 
       &  &  & $\alpha$ & 0.60 & 0.0247 & 0.954 &  & 0.0530 & 0.930 \\* 
       &  &  & $\eta$ & 5.00 & -0.0149 & 0.956 &  & 0.4039 & 0.972 \\* 
       &  &  & $\delta$ & 0.55 & 0.0238 & 0.972 &  & 0.0567 & 0.946 \\ 
      10 & 0.55 & 25 & $\beta_0$ & 2.00 & -0.0502 & 0.952 &  & -0.0278 & 0.976 \\* 
       &  &  & $\beta_1$ & 0.30 & -0.0153 & 0.944 &  & 0.0075 & 0.968 \\* 
       &  &  & $\alpha$ & 0.60 & 0.3571 & 0.892 &  & 1.3843 & 0.858 \\* 
       &  &  & $\eta$ & 10.00 & 0.2566 & 0.988 &  & 18.6893 & 0.984 \\* 
       &  &  & $\delta$ & 0.55 & 0.1092 & 0.970 &  & 0.1691 & 0.992 \\ 
      10 & 0.55 & 50 & $\beta_0$ & 2.00 & -0.0513 & 0.946 &  & -0.0406 & 0.950 \\* 
       &  &  & $\beta_1$ & 0.30 & -0.0106 & 0.964 &  & 0.0064 & 0.964 \\* 
       &  &  & $\alpha$ & 0.60 & 0.1152 & 0.920 &  & 0.3609 & 0.888 \\* 
       &  &  & $\eta$ & 10.00 & -0.4251 & 0.952 &  & 17.5202 & 0.964 \\* 
       &  &  & $\delta$ & 0.55 & 0.0564 & 0.974 &  & 0.1215 & 0.966 \\ 
      10 & 0.55 & 100 & $\beta_0$ & 2.00 & -0.0250 & 0.942 &  & -0.0240 & 0.934 \\* 
       &  &  & $\beta_1$ & 0.30 & 0.0003 & 0.948 &  & -0.0025 & 0.938 \\* 
       &  &  & $\alpha$ & 0.60 & 0.0349 & 0.924 &  & 0.1134 & 0.894 \\* 
       &  &  & $\eta$ & 10.00 & -0.1416 & 0.936 &  & 11.5311 & 0.940 \\* 
       &  &  & $\delta$ & 0.55 & 0.0293 & 0.972 &  & 0.0742 & 0.968 \\ 
      10 & 0.55 & 200 & $\beta_0$ & 2.00 & -0.0136 & 0.940 &  & -0.0172 & 0.960 \\* 
       &  &  & $\beta_1$ & 0.30 & -0.0045 & 0.966 &  & -0.0016 & 0.952 \\* 
       &  &  & $\alpha$ & 0.60 & 0.0211 & 0.944 &  & 0.0476 & 0.932 \\* 
       &  &  & $\eta$ & 10.00 & -0.0728 & 0.962 &  & 2.9262 & 0.952 \\* 
       &  &  & $\delta$ & 0.55 & 0.0145 & 0.970 &  & 0.0399 & 0.972 \\ 
    
    \end{longtable}

    \afterpage{
    
        \begin{landscape}
        
            \pagestyle{landscape}
            
            \begin{longtable}{lS[table-format=2.0]S[table-format=1.2]S[table-format=3.0]lS[table-format=1.2]S[table-format=2.4]S[table-format=1.4]S[table-format=1.3]S[table-format=1.4]lS[table-format=2.4]S[table-format=1.4]S[table-format=1.3]S[table-format=1.4]}
                \caption{Simulation results for the TDW and cTDW models across all scenarios. For each combination of $c$, $\eta$, $\delta$, and sample size $N$, 500 datasets were generated. Here, $\delta$ denotes the weight of the narrower component in the cTDW model. The table reports the mean bias, RMSE, 95\% BCI coverage probability (CP), and average 95\% BCI length for the regression coefficients.}
                \label{tab:sim_result_beta} \\
                \hline
                & & & & & & \multicolumn{9}{c}{\bf Model} \\
                \cline{7-15}
                & & & & & & \multicolumn{4}{c}{\bf TDW} & & \multicolumn{4}{c}{\bf cTDW} \\
                \cline{7-10} \cline{12-15}
                $\bm c$ & $\bm\eta$ & $\bm\delta$ & {$\bm N$} & {\bf Parameter} & {\bf Value} & {\bf Bias} & {\bf RMSE} & {\bf CP} & {\bf BCI lgth} & & {\bf Bias} & {\bf RMSE} & {\bf CP} & {\bf BCI lgth} \\
                \hline
                \endfirsthead
              
                \multicolumn{15}{c}
                {{\bfseries \tablename\ \thetable{} -- continued from previous page}} \\
                \hline
                & & & & & & \multicolumn{9}{c}{\bf Model} \\
                \cline{7-15}
                & & & & & & \multicolumn{4}{c}{\bf TDW} & & \multicolumn{4}{c}{\bf cTDW} \\
                \cline{7-10} \cline{12-15}
                $\bm c$ & $\bm\eta$ & $\bm\delta$ & {$\bm N$} & {\bf Parameter} & {\bf Value} & {\bf Bias} & {\bf RMSE} & {\bf CP} & {\bf BCI lgth} & & {\bf Bias} & {\bf RMSE} & {\bf CP} & {\bf BCI lgth} \\
                \hline
                \endhead
              
                \hline \multicolumn{15}{r}{{Continued on next page}} \\
                \endfoot
              
                \hline
                \endlastfoot

                  0 & 2 & 0.9 & 25 & $\beta_0$ & 2.00 & -0.0146 & 0.1621 & 0.942 & 0.6735 &  & 0.0014 & 0.1600 & 0.936 & 0.6330 \\* 
                   &  &  &  & $\beta_1$ & 0.30 & 0.0056 & 0.1566 & 0.952 & 0.6254 &  & 0.0036 & 0.1503 & 0.950 & 0.6109 \\* 
                  0 & 2 & 0.9 & 50 & $\beta_0$ & 2.00 & -0.0028 & 0.1097 & 0.964 & 0.4690 &  & 0.0067 & 0.1093 & 0.944 & 0.4314 \\* 
                   &  &  &  & $\beta_1$ & 0.30 & -0.0038 & 0.1153 & 0.948 & 0.4183 &  & -0.0008 & 0.1046 & 0.946 & 0.4007 \\* 
                  0 & 2 & 0.9 & 100 & $\beta_0$ & 2.00 & -0.0076 & 0.0747 & 0.974 & 0.3318 &  & 0.0021 & 0.0752 & 0.954 & 0.3065 \\* 
                   &  &  &  & $\beta_1$ & 0.30 & 0.0062 & 0.0885 & 0.906 & 0.2884 &  & 0.0032 & 0.0714 & 0.954 & 0.2778 \\* 
                  0 & 2 & 0.9 & 200 & $\beta_0$ & 2.00 & -0.0138 & 0.0555 & 0.974 & 0.2339 &  & -0.0041 & 0.0534 & 0.960 & 0.2154 \\ 
                   &  &  &  & $\beta_1$ & 0.30 & 0.0003 & 0.0605 & 0.902 & 0.2023 &  & -0.0011 & 0.0492 & 0.952 & 0.1941 \\* 
                  0 & 5 & 0.55 & 25 & $\beta_0$ & 2.00 & 0.0871 & 0.4384 & 0.964 & 1.9925 &  & 0.0146 & 0.3152 & 0.956 & 1.3559 \\* 
                   &  &  &  & $\beta_1$ & 0.30 & -0.0021 & 0.5375 & 0.890 & 1.7950 &  & 0.0008 & 0.3643 & 0.936 & 1.3660 \\* 
                  0 & 5 & 0.55 & 50 & $\beta_0$ & 2.00 & 0.0668 & 0.3015 & 0.978 & 1.4335 &  & -0.0176 & 0.1820 & 0.944 & 0.7900 \\* 
                   &  &  &  & $\beta_1$ & 0.30 & -0.0155 & 0.4036 & 0.874 & 1.2630 &  & -0.0071 & 0.1986 & 0.950 & 0.7758 \\* 
                  0 & 5 & 0.55 & 100 & $\beta_0$ & 2.00 & 0.0523 & 0.2262 & 0.976 & 1.0086 &  & -0.0148 & 0.1225 & 0.942 & 0.5003 \\* 
                   &  &  &  & $\beta_1$ & 0.30 & -0.0129 & 0.3134 & 0.824 & 0.8736 &  & -0.0066 & 0.1218 & 0.934 & 0.4616 \\* 
                  0 & 5 & 0.55 & 200 & $\beta_0$ & 2.00 & 0.0424 & 0.1645 & 0.972 & 0.7144 &  & -0.0104 & 0.0909 & 0.940 & 0.3410 \\* 
                   &  &  &  & $\beta_1$ & 0.30 & 0.0056 & 0.2067 & 0.868 & 0.6108 &  & -0.0010 & 0.0730 & 0.958 & 0.3058 \\ 
                  0 & 10 & 0.55 & 25 & $\beta_0$ & 2.00 & 0.2571 & 0.8224 & 0.974 & 3.8480 &  & -0.0502 & 0.3804 & 0.952 & 1.6178 \\* 
                   &  &  &  & $\beta_1$ & 0.30 & -0.0098 & 1.0845 & 0.872 & 3.4634 &  & -0.0153 & 0.4652 & 0.944 & 1.6750 \\* 
                  0 & 10 & 0.55 & 50 & $\beta_0$ & 2.00 & 0.2038 & 0.5987 & 0.980 & 2.7492 &  & -0.0513 & 0.2084 & 0.946 & 0.8289 \\* 
                   &  &  &  & $\beta_1$ & 0.30 & -0.0342 & 0.7967 & 0.854 & 2.3984 &  & -0.0106 & 0.1860 & 0.964 & 0.7765 \\* 
                  0 & 10 & 0.55 & 100 & $\beta_0$ & 2.00 & 0.1940 & 0.4386 & 0.978 & 1.9536 &  & -0.0250 & 0.1328 & 0.942 & 0.5043 \\* 
                   &  &  &  & $\beta_1$ & 0.30 & -0.0212 & 0.5745 & 0.824 & 1.6695 &  & 0.0003 & 0.1160 & 0.948 & 0.4413 \\* 
                  0 & 10 & 0.55 & 200 & $\beta_0$ & 2.00 & 0.1804 & 0.3181 & 0.980 & 1.4075 &  & -0.0136 & 0.0922 & 0.940 & 0.3421 \\* 
                   &  &  &  & $\beta_1$ & 0.30 & 0.0099 & 0.4140 & 0.858 & 1.1889 &  & -0.0045 & 0.0722 & 0.966 & 0.2900 \\ 
                  1 & 2 & 0.9 & 25 & $\beta_0$ & 2.00 & -0.0247 & 0.1830 & 0.952 & 0.7438 &  & -0.0075 & 0.1849 & 0.948 & 0.7309 \\* 
                   &  &  &  & $\beta_1$ & 0.30 & 0.0051 & 0.1687 & 0.960 & 0.6775 &  & 0.0089 & 0.1696 & 0.964 & 0.6882 \\* 
                  1 & 2 & 0.9 & 50 & $\beta_0$ & 2.00 & -0.0158 & 0.1187 & 0.956 & 0.5114 &  & -0.0028 & 0.1175 & 0.944 & 0.4955 \\* 
                   &  &  &  & $\beta_1$ & 0.30 & -0.0000 & 0.1200 & 0.926 & 0.4411 &  & 0.0019 & 0.1132 & 0.952 & 0.4441 \\* 
                  1 & 2 & 0.9 & 100 & $\beta_0$ & 2.00 & -0.0202 & 0.0893 & 0.956 & 0.3593 &  & -0.0070 & 0.0861 & 0.950 & 0.3409 \\* 
                   &  &  &  & $\beta_1$ & 0.30 & 0.0022 & 0.0899 & 0.926 & 0.3084 &  & 0.0029 & 0.0789 & 0.942 & 0.3029 \\* 
                  1 & 2 & 0.9 & 200 & $\beta_0$ & 2.00 & -0.0114 & 0.0649 & 0.946 & 0.2518 &  & 0.0013 & 0.0619 & 0.934 & 0.2361 \\* 
                   &  &  &  & $\beta_1$ & 0.30 & 0.0012 & 0.0593 & 0.926 & 0.2125 &  & 0.0012 & 0.0534 & 0.946 & 0.2082 \\ 
                  1 & 5 & 0.55 & 25 & $\beta_0$ & 2.00 & -0.0454 & 0.2986 & 0.974 & 1.3798 &  & -0.0314 & 0.2657 & 0.966 & 1.1904 \\* 
                   &  &  &  & $\beta_1$ & 0.30 & 0.0137 & 0.2979 & 0.966 & 1.2643 &  & -0.0130 & 0.2570 & 0.962 & 1.1455 \\* 
                  1 & 5 & 0.55 & 50 & $\beta_0$ & 2.00 & -0.0749 & 0.2324 & 0.956 & 0.9700 &  & -0.0343 & 0.1929 & 0.934 & 0.7575 \\* 
                   &  &  &  & $\beta_1$ & 0.30 & -0.0228 & 0.2326 & 0.932 & 0.8383 &  & -0.0166 & 0.1811 & 0.948 & 0.7092 \\* 
                  1 & 5 & 0.55 & 100 & $\beta_0$ & 2.00 & -0.0773 & 0.1665 & 0.970 & 0.6969 &  & -0.0138 & 0.1246 & 0.952 & 0.5005 \\* 
                   &  &  &  & $\beta_1$ & 0.30 & -0.0074 & 0.1739 & 0.908 & 0.5905 &  & -0.0101 & 0.1247 & 0.934 & 0.4665 \\* 
                  1 & 5 & 0.55 & 200 & $\beta_0$ & 2.00 & -0.0849 & 0.1353 & 0.930 & 0.4902 &  & -0.0039 & 0.0835 & 0.948 & 0.3371 \\* 
                   &  &  &  & $\beta_1$ & 0.30 & 0.0095 & 0.1160 & 0.896 & 0.4119 &  & 0.0000 & 0.0815 & 0.938 & 0.3120 \\ 
                  1 & 10 & 0.55 & 25 & $\beta_0$ & 2.00 & -0.0576 & 0.3165 & 0.984 & 1.6177 &  & -0.0278 & 0.2686 & 0.976 & 1.2762 \\* 
                   &  &  &  & $\beta_1$ & 0.30 & 0.0168 & 0.3853 & 0.948 & 1.4770 &  & 0.0075 & 0.2981 & 0.968 & 1.2470 \\* 
                  1 & 10 & 0.55 & 50 & $\beta_0$ & 2.00 & -0.1210 & 0.2643 & 0.976 & 1.1631 &  & -0.0406 & 0.1922 & 0.950 & 0.8079 \\* 
                   &  &  &  & $\beta_1$ & 0.30 & 0.0053 & 0.2692 & 0.932 & 1.0146 &  & 0.0064 & 0.1842 & 0.964 & 0.7732 \\* 
                  1 & 10 & 0.55 & 100 & $\beta_0$ & 2.00 & -0.1476 & 0.2274 & 0.952 & 0.8298 &  & -0.0240 & 0.1315 & 0.934 & 0.5144 \\* 
                   &  &  &  & $\beta_1$ & 0.30 & -0.0049 & 0.1988 & 0.912 & 0.7072 &  & -0.0025 & 0.1263 & 0.938 & 0.4805 \\* 
                  1 & 10 & 0.55 & 200 & $\beta_0$ & 2.00 & -0.1635 & 0.2012 & 0.866 & 0.5854 &  & -0.0172 & 0.0921 & 0.960 & 0.3541 \\* 
                   &  &  &  & $\beta_1$ & 0.30 & 0.0036 & 0.1361 & 0.926 & 0.4963 &  & -0.0016 & 0.0830 & 0.952 & 0.3226 \\ 
    
            \end{longtable}
            
        \end{landscape}
    }

    \section{Discussion} \label{sec:conclusions}

    In this paper, we introduce a cTDW framework for modeling robust count data. Building on the TDW distribution, we introduce a heavier-tailed subcomponent that absorbs outliers while preserving a single median parameter $m^*$. Applying the cTDW model to Arizona hospital length-of-stay data (truncated at $c = 1$) yields marked improvements in residual diagnostics and predictive accuracy (via LOO cross-validation) relative to the single-component TDW model, particularly in capturing extreme hospital stays. Moreover, K-L divergence checks show that the cTDW approach flags far fewer high-influence points than the TDW model, indicating its capacity to handle outliers effectively.

    Two interpretations of the cTDW model are possible. Under the classical contamination view, the narrower component represents the bulk of the data, while the heavier-tailed component captures a minority of the data that deviates from that bulk. Under a more general heavy-tailed mixture view, the two-component model is used to accommodate tail behavior without requiring the heavier-tailed component to be a minority component. Our primary analysis reflects the former interpretation through the prior restriction on $\delta$, whereas the alternative prior specification in the sensitivity analysis is more consistent with the latter. In the Arizona application, both specifications gave very similar posterior results, so the substantive conclusions do not depend materially on which interpretation is emphasized. Throughout, the term ``contaminated'' refers to the addition of a heavier-tailed component to a baseline TDW while retaining a common shifted median.

    For completeness, we fit a TNB solely as a predictive benchmark. Its predictive performance was intermediate (worse than cTDW, better than TDW); the K-L divergence profile was considerably milder than TDW, but the residual QQ plot still showed departures from uniformity. Given the lack of a clear coefficient interpretation under truncation, we did not interpret the TNB parameters.

    Our simulation study compared the cTDW model with a single-component TDW model across a range of mixture settings, sample sizes, and lower bounds. Under the mild mixture setting, the two models performed similarly, as expected when the heavier-tail component carries relatively little weight. Under the more pronounced mixture settings, however, the TDW model exhibited inflated RMSE, wider BCIs, and poorer coverage, whereas the cTDW model remained considerably more stable. These gains were most evident for the regression coefficients when $c = 0$, with similar but somewhat less pronounced improvements under truncation at $c = 1$. The simulation results also showed that the tail-mixture parameters, especially $\eta$, can be harder to estimate in smaller samples and under truncation. Overall, these findings suggest that the heavier-tail subcomponent improves robustness when the data arise from a genuinely heavy-tailed mixture.

    Although we focus on $c=1$ for strictly positive data (as in the length of stay), setting $c=0$ allows the cTDW model to cover the full range $\left\{0,1,2,\dots\right\}$. A heavier-tailed component, however, can increase mass at both extremes. When genuinely large values drive overdispersion, but the data do not contain extra zeros, this left-tail behavior can spuriously inflate the fitted probability at $Y=0$. The same issue is known for heavy-tailed Poisson mixtures~\citep{karlis2005mixed}, such as the generalized gamma or Sichel models. A practical remedy is to separate the zero cell from the positive counts by embedding the cTDW in a hurdle framework: the hurdle models $P\left(Y=0\right)$ directly, while the heavy-tail mixture is applied only to $Y\in\left\{1,2,\dots\right\}$. Concretely, one may define
    \begin{equation}
        P\left(Y = y\right) =
        \begin{cases}
            \pi, & y = 0, \\
            \left(1 - \pi\right) P_{\mathrm{cTDW}}\left(Y = y \left|m^*, \alpha, \eta, \delta, c = 1\right.\right), & y \in \left\{1,2,\dots \right\},
        \end{cases}
    \end{equation}
    thereby avoiding over-predicting zeros or misrepresenting heavy right tails among the positive counts.

    Overall, our results suggest that the cTDW model provides a flexible and robust framework for analyzing count data with outliers, particularly when the median is the preferred measure of location and a heavier-tailed mixture component is needed to capture the observed upper tail. This conclusion was unchanged under both the classical contamination interpretation and the broader heavy-tailed mixture interpretation considered in the sensitivity analysis. The contaminated structure provides a practical alternative to conventional overdispersed count models in settings where extremely large counts drive the overdispersion.
    
    \section*{Research ethics}

    Not applicable.

    \section*{Author contributions}

    All authors have accepted responsibility for the entire content of this manuscript and approved its submission.
        
    \section*{Competing interests}
        
    The authors state no competing interests.

    \section*{Research funding}

    This work is based on the research supported by the National Research Foundation (NRF) of South Africa (Grant number 132383). Opinions expressed and conclusions arrived at are those of the authors and are not necessarily to be attributed to the NRF.

    \section*{Data availability statement}
    
    The Arizona hospital LOS dataset used in this article is freely available from the \texttt{COUNT} package in R. All R code for reproducing the application results can be found on GitHub at \href{https://github.com/DABURGER1/Robust-Count-cTDW}{Robust-Count-cTDW}.

    \begin{appendices}

    \renewcommand{\thesection}{Appendix \Alph{section}}
    \renewcommand{\thesubsection}{\Alph{section}.\arabic{subsection}}
    \renewcommand{\theequation}{\Alph{section}.\arabic{equation}}
    \setcounter{subsection}{0}
    \setcounter{equation}{0}

    \section{Pearson kurtosis of truncated discrete Weibull distributions} \label{sec:DW_dists_kurtosis}
    
    \subsection{Truncated discrete Weibull kurtosis}\label{sec:TDW_kurtosis}
    
    To state the moment formulas compactly, let $a > 0$ denote a generic TDW dispersion parameter. Under the $\left(m^*, a\right)$ parameterization,
    \begin{equation}
        \label{eq:q_a_def}
        q_a
        =
        \exp\left[
        \frac{\ln\left(0.5\right)}
        {\left(m^*\right)^{1/a} - c^{1/a}}
        \right].
    \end{equation}
    Then, for $Y \sim \mathrm{TDW}\left(m^*, a, c\right)$,
    \begin{equation}
        P\left(Y = y \middle| m^*, a, c\right)
        =
        \frac{q_a^{y^{1/a}} - q_a^{\left(y+1\right)^{1/a}}}{q_a^{c^{1/a}}},
        \qquad y = c,c+1,\ldots .
    \end{equation}
    The $k$\textsuperscript{th} raw moment of $Y$, for an integer $k \ge 1$, is therefore
    \begin{equation}
        \label{eq:TDW_raw_moment}
        \mathbb{E}\left(Y^k \middle| m^*, a, c\right)
        =
        \sum\limits_{y=c}^{\infty} y^k P\left(Y = y \middle| m^*, a, c\right)
        =
        \frac{1}{q_a^{c^{1/a}}}
        \sum\limits_{y=c}^{\infty}
        y^k
        \left(
        q_a^{y^{1/a}} - q_a^{\left(y+1\right)^{1/a}}
        \right).
    \end{equation}
    We use the finite-difference identity
    \begin{equation}
        \label{eq:diff_of_powers}
        n^k - \left(n-1\right)^k
        =
        \sum\limits_{j=0}^{k-1}
        \binom{k}{j}
        \left(-1\right)^{k-1-j}
        n^j,
    \end{equation}
    which implies
    \begin{equation}
        \label{eq:nk_expand}
        n^k
        =
        \sum\limits_{m=1}^{n}
        \left(
        m^k - \left(m-1\right)^k
        \right).
    \end{equation}
    Hence,
    \begin{equation}
        y^k
        =
        \sum\limits_{m=1}^{y}
        \left(
        m^k - \left(m-1\right)^k
        \right).
    \end{equation}
    Substituting into \eqref{eq:TDW_raw_moment} and applying Tonelli's theorem \citep{folland1999real} gives
    \begin{equation}
        \mathbb{E}\left(Y^k \middle| m^*, a, c\right)
        =
        \frac{1}{q_a^{c^{1/a}}}
        \sum\limits_{m=1}^{\infty}
        \left(
        m^k - \left(m-1\right)^k
        \right)
        \sum\limits_{y=\max\left(c,m\right)}^{\infty}
        \left(
        q_a^{y^{1/a}} - q_a^{\left(y+1\right)^{1/a}}
        \right).
    \end{equation}
    If $1 \le m \le c$, the inner sum starts at $y=c$; if $m \ge c+1$, it starts at $y=m$. After telescoping,
    \begin{equation}
        \sum\limits_{y=m}^{\infty}
        \left(
        q_a^{y^{1/a}} - q_a^{\left(y+1\right)^{1/a}}
        \right)
        =
        q_a^{m^{1/a}},
        \qquad
        \sum\limits_{y=c}^{\infty}
        \left(
        q_a^{y^{1/a}} - q_a^{\left(y+1\right)^{1/a}}
        \right)
        =
        q_a^{c^{1/a}}.
    \end{equation}
    Here, we use $0 < q_a < 1$ and $a > 0$, so $q_a^{\left(N+1\right)^{1/a}} \to 0$ as $N \to \infty$. Because $0 < q_a < 1$, the terms $q_a^{m^{1/a}}$ decay to zero as $m \to \infty$. Therefore, for any integer $k \ge 1$, the series defining $\mathbb{E}\left(Y^k \middle| m^*, a, c\right)$ converges, so the raw moments of all orders exist.
    
    Combining terms yields
    \begin{equation}
        \label{eq:TDW_moment_general_c}
        \mathbb{E}\left(Y^k \middle| m^*, a, c\right)
        =
        c^k
        +
        \frac{1}{q_a^{c^{1/a}}}
        \sum\limits_{m=c+1}^{\infty}
        \left(
        m^k - \left(m-1\right)^k
        \right)
        q_a^{m^{1/a}},
    \end{equation}
    which is valid for all $c \in \left\{0,1,2,\ldots\right\}$ and integers $k \ge 1$.
    
    If $c=1$, then $q_a^{c^{1/a}} = q_a$ and $c^k = 1$, so \eqref{eq:TDW_moment_general_c} simplifies to
    \begin{equation}
        \label{eq:TDW_moment_c1}
        \mathbb{E}\left(Y^k \middle| m^*, a, 1\right)
        =
        \frac{1}{q_a}
        \sum\limits_{m=1}^{\infty}
        \left(
        m^k - \left(m-1\right)^k
        \right)
        q_a^{m^{1/a}}.
    \end{equation}
    For notational convenience, define
    \begin{equation}
        \label{eq:TDW_Mk_def}
        M_k^{\mathrm{TDW}}\left(a\right)
        =
        \mathbb{E}\left(Y^k \middle| m^*, a, c\right),
        \qquad k = 1,2,3,4.
    \end{equation}
    Then the mean and variance of the TDW distribution are
    \begin{equation}
        \mu_{\mathrm{TDW}} = M_1^{\mathrm{TDW}}\left(a\right),
        \qquad
        \sigma_{\mathrm{TDW}}^2
        =
        M_2^{\mathrm{TDW}}\left(a\right) - \left(M_1^{\mathrm{TDW}}\left(a\right)\right)^2,
    \end{equation}
    and the fourth central moment is
    \begin{equation}
        \mu_{4,\mathrm{TDW}}
        =
        M_4^{\mathrm{TDW}}\left(a\right)
        - 4 M_1^{\mathrm{TDW}}\left(a\right) M_3^{\mathrm{TDW}}\left(a\right)
        + 6 \left(M_1^{\mathrm{TDW}}\left(a\right)\right)^2 M_2^{\mathrm{TDW}}\left(a\right)
        - 3 \left(M_1^{\mathrm{TDW}}\left(a\right)\right)^4.
    \end{equation}
    Hence, the Pearson kurtosis of the TDW distribution is
    \begin{equation}
        \label{eq:TDW_KURT}
        \mathrm{Kurt}_{\mathrm{TDW}}\left(m^*, a, c\right)
        =
        \frac{\mu_{4,\mathrm{TDW}}}{\left(\sigma_{\mathrm{TDW}}^2\right)^2}.
    \end{equation}
    In the single-component TDW model considered in the main text, one simply sets $a = \alpha$.
    
    \subsection{Contaminated truncated discrete Weibull kurtosis}\label{sec:CTDW_kurtosis}
    
    Now consider
    \begin{equation}
        \label{eq:CTDW_def_kurt}
        Y
        \sim
        \delta \, \mathrm{TDW}\left(m^*, \alpha, c\right)
        +
        \left(1 - \delta\right)\mathrm{TDW}\left(m^*, \eta\alpha, c\right),
    \end{equation}
    where $0 < \delta < 1$ and $\eta > 1$. Let
    \begin{equation}
        Y_1 \sim \mathrm{TDW}\left(m^*, \alpha, c\right),
        \qquad
        Y_2 \sim \mathrm{TDW}\left(m^*, \eta\alpha, c\right).
    \end{equation}
    Since the raw moments of a finite mixture are weighted averages of the component raw moments, it follows that
    \begin{equation}
        \label{eq:CTDW_raw_moment_mix}
        \mathbb{E}\left(Y^k \middle| m^*, \alpha, \eta, \delta, c\right)
        =
        \delta \mathbb{E}\left(Y_1^k\right)
        +
        \left(1 - \delta\right)\mathbb{E}\left(Y_2^k\right),
        \qquad k = 1,2,3,4.
    \end{equation}
    Using the TDW raw-moment function in \eqref{eq:TDW_Mk_def}, define
    \begin{equation}
        M_k^{\mathrm{cTDW}}
        =
        \mathbb{E}\left(Y^k \middle| m^*, \alpha, \eta, \delta, c\right).
    \end{equation}
    Then
    \begin{equation}
        \label{eq:CTDW_Mk_compact}
        M_k^{\mathrm{cTDW}}
        =
        \delta M_k^{\mathrm{TDW}}\left(\alpha\right)
        +
        \left(1 - \delta\right)M_k^{\mathrm{TDW}}\left(\eta\alpha\right),
        \qquad k = 1,2,3,4.
    \end{equation}
    Equivalently, substituting \eqref{eq:TDW_moment_general_c} gives
    \begin{equation}
        \label{eq:CTDW_Mk_expanded}
        M_k^{\mathrm{cTDW}}
        =
        c^k
        +
        \delta
        \frac{1}{q_{\alpha}^{c^{1/\alpha}}}
        \sum\limits_{m=c+1}^{\infty}
        \left(
        m^k - \left(m-1\right)^k
        \right)
        q_{\alpha}^{m^{1/\alpha}}
        +
        \left(1 - \delta\right)
        \frac{1}{q_{\eta\alpha}^{c^{1/\left(\eta\alpha\right)}}}
        \sum\limits_{m=c+1}^{\infty}
        \left(
        m^k - \left(m-1\right)^k
        \right)
        q_{\eta\alpha}^{m^{1/\left(\eta\alpha\right)}},
    \end{equation}
    where
    \begin{equation}
        q_{\alpha}
        =
        \exp\left[
        \frac{\ln\left(0.5\right)}
        {\left(m^*\right)^{1/\alpha} - c^{1/\alpha}}
        \right],
        \qquad
        q_{\eta\alpha}
        =
        \exp\left[
        \frac{\ln\left(0.5\right)}
        {\left(m^*\right)^{1/\left(\eta\alpha\right)} - c^{1/\left(\eta\alpha\right)}}
        \right].
    \end{equation}
    That is, $q_{\alpha}$ and $q_{\eta\alpha}$ are obtained from \eqref{eq:q_a_def} by setting $a = \alpha$ and $a = \eta\alpha$, respectively.
    
    When $c=1$, \eqref{eq:CTDW_Mk_expanded} simplifies to
    \begin{equation}
        \label{eq:CTDW_Mk_c1}
        M_k^{\mathrm{cTDW}}
        =
        \delta
        \frac{1}{q_{\alpha}}
        \sum\limits_{m=1}^{\infty}
        \left(
        m^k - \left(m-1\right)^k
        \right)
        q_{\alpha}^{m^{1/\alpha}}
        +
        \left(1 - \delta\right)
        \frac{1}{q_{\eta\alpha}}
        \sum\limits_{m=1}^{\infty}
        \left(
        m^k - \left(m-1\right)^k
        \right)
        q_{\eta\alpha}^{m^{1/\left(\eta\alpha\right)}}.
    \end{equation}
    Using the mixed raw moments $M_k^{\mathrm{cTDW}}$, $k = 1,2,3,4$, the Pearson kurtosis of the cTDW distribution is obtained exactly as in the TDW case, with $M_k^{\mathrm{TDW}}\left(a\right)$ replaced by $M_k^{\mathrm{cTDW}}$. In particular,
    \begin{equation}
        \mu_{\mathrm{cTDW}} = M_1^{\mathrm{cTDW}},
        \qquad
        \sigma_{\mathrm{cTDW}}^2 = M_2^{\mathrm{cTDW}} - \left(M_1^{\mathrm{cTDW}}\right)^2,
    \end{equation}
    \begin{equation}
        \mu_{4,\mathrm{cTDW}}
        =
        M_4^{\mathrm{cTDW}}
        - 4 M_1^{\mathrm{cTDW}} M_3^{\mathrm{cTDW}}
        + 6 \left(M_1^{\mathrm{cTDW}}\right)^2 M_2^{\mathrm{cTDW}}
        - 3 \left(M_1^{\mathrm{cTDW}}\right)^4,
    \end{equation}
    and hence
    \begin{equation}
        \label{eq:CTDW_KURT}
        \mathrm{Kurt}_{\mathrm{cTDW}}\left(m^*, \alpha, \eta, \delta, c\right)
        =
        \frac{\mu_{4,\mathrm{cTDW}}}{\left(\sigma_{\mathrm{cTDW}}^2\right)^2}.
    \end{equation}
    
    \section{Kullback-Leibler divergence for outlier detection}
    \label{sec:KL_METHOD}
        
    To assess the influence of each individual observation on the TDW or cTDW fit, we adopt a LOO approach similar to that of Wang and Luo \citep{WANG2016}. Specifically, we evaluate the K-L divergence between the posterior distributions obtained with and without a given observation. Large K-L values indicate that omitting a certain data point meaningfully shifts the posterior, suggesting that the point may be an outlier.
        
    Denote by $\bm{\theta}^{\left(k\right)}$ the $k$\textsuperscript{th} posterior draw of these parameters, for $k = 1, \dots, K$. Suppose the dataset $\mathcal{D}$ contains observations $\mathcal{D}_i$. We write $P\left(\bm{\theta} \left|\mathcal{D}\right.\right)$ for the posterior given the full dataset and $P\left(\bm{\theta} \left|\mathcal{D}_{\left[i\right]}\right.\right)$ for the posterior distribution when observation $\mathcal{D}_i$ is removed. Following Wang \& Luo \citep{WANG2016}, the K-L divergence between these two posteriors under model $R$ (TDW or cTDW) is approximated as:
    \begin{equation}
        \label{eq:KL_approx}
            \mathrm{KL}_R\left(P\left(\bm{\theta} \left|\mathcal{D}\right.\right), P\left(\bm{\theta} \left|\mathcal{D}_{\left[i\right]}\right.\right)\right)
            = \log \left\{\frac{1}{K} \sum\limits_{k=1}^{K} \left[P\left(\mathcal{D}_i \left|\bm{\theta}^{\left(k\right)}\right.\right)\right]^{-1} \right\}
            + \frac{1}{K} \sum\limits_{k=1}^{K} \log \left[P\left(\mathcal{D}_i \left|\bm{\theta}^{\left(k\right)}\right.\right)\right].
    \end{equation}
    In \eqref{eq:KL_approx}, each $P\left(\mathcal{D}_i \left|\bm{\theta}^{\left(k\right)}\right.\right)$ corresponds to the model likelihood (TDW or cTDW) evaluated at the posterior draw $\bm{\theta}^{\left(k\right)}$. A higher $\mathrm{KL}_R$ value indicates that removing $\mathcal{D}_i$ significantly alters the posterior, implying that the point may be influential or an outlier under model $R$.
        
    We follow Tomazella et al. \citep{TOMAZELLA2021} in labeling an observation $\mathcal{D}_i$ as an outlier if:
    \begin{equation}
        \label{eq:flag_outlier}
            0.5 \left(1 + \sqrt{ 1 - \exp \left[-2 \mathrm{KL}_R\left(P\left(\bm{\theta} \left|\mathcal{D}\right.\right), P\left(\bm{\theta} \left|\mathcal{D}_{\left[i\right]}\right.\right)\right)\right] }\right) \geq 0.8.
    \end{equation}
    This criterion flags data points that, when omitted, significantly alter the posterior distribution, indicating an unusually large influence on parameter estimates. In practice, $\mathrm{KL}_R$ is computed for each $\mathcal{D}_i$, and \eqref{eq:flag_outlier} is applied to identify points that have a disproportionate impact on the posterior. For the cTDW model, posterior component membership probabilities can also serve as a complementary observation-level summary in Bayesian contaminated models. \citep{okhli2021contaminated}. These probabilities measure how strongly each observation is associated with the heavier-tail component, whereas the K-L divergence measures how much posterior inference changes when that observation is omitted. We focus on the K-L diagnostic because it provides a unified measure of case-deletion influence across all fitted models. Moreover, if a binary outlier classification is desired, posterior component membership probabilities also require a cutoff, whereas the calibration used here follows Tomazella et al. \citep{TOMAZELLA2021}.

    \end{appendices}

    \bibliography{Bibliography}

@article{BURGER2021,
    Title = {{Nonlinear mixed-effects modeling of longitudinal count data: Bayesian inference about median counts based on the marginal zero-inflated discrete Weibull distribution}},
    Author = {Burger, D. A. and Lesaffre, E.},
    Journal = {Statistics in Medicine},
    doi = {https://doi.org/10.1002/sim.9112},
    Volume = {40},
    Number = {23},
    Pages = {5078-5095},
    Year = {2021}}

@article{WANG2016,
    Title = {{Augmented Beta rectangular regression models: A Bayesian perspective}},
    Author = {Wang, J. and Luo, S.},
    Journal = {Biometrical Journal},
    Volume = {58},
    Number = {1},
    Pages = {206--221},
    doi = {https://doi.org/10.1002/bimj.201400232},
    Year = {2016}}

@article{BROOKS1998,
    Author = {Brooks, S. P. and Gelman, A.},
    Journal = {Journal of Computational and Graphical Statistics},
    Number = {4},
    Pages = {434--455},
    doi = {https://doi.org/10.1080/10618600.1998.10474787},
    Title = {{General methods for monitoring convergence of iterative simulations}},
    Volume = {7},
    Year = {1998}}

@article{TOMAZELLA2021,
    Title = {{Bayesian reference analysis for the generalized normal linear regression model}},
    Author = {Tomazella, V. L. D. and Jesus, S. and Gazon, A. B. and Louzada, F. and Nadarajah, S. and Nascimento, D. C. and Rodrigues, F. A. and Ramos, P. L.},
    Journal = {Symmetry},
    Volume = {13},
    Number = {5},
    Pages = {856},
    doi={https://doi.org/10.3390/sym13050856},
    Year = {2021}}

@article{DENWOOD2016,
    Title = {{runjags: An R package providing interface utilities, model templates, parallel computing methods and additional distributions for MCMC models in JAGS}},
    Author = {Denwood, M. J.},
    Journal = {Journal of Statistical Software},
    doi = {https://doi.org/10.18637/jss.v071.i09},
    Volume = {71},
    Number = {9},
    Pages = {1--25},
    Year = {2016}}

@article{burger2020robust,
  title={{A robust Bayesian mixed effects approach for zero inflated and highly skewed longitudinal count data emanating from the zero inflated discrete Weibull distribution}},
  author={Burger, D. A. and Schall, R. and Ferreira, J. T. and Chen, D.-G.},
  doi={https://doi.org/10.1002/sim.8475},
  journal={Statistics in Medicine},
  volume={39},
  number={9},
  pages={1275--1291},
  year={2020}}

@manual{hartig2024package,
    Title = {{DHARMa: residual diagnostics for hierarchical (multi-level / mixed)
regression models}},
    Author = {Hartig, F. and Lohse, L. and {de Souza}, M.},
    Year = {2024},
    Note = {{R package version 0.4.7}},
    Url = {https://cran.r-project.org/web/packages/DHARMa/index.html}}

@manual{vehtari2024package,
    Title = {{loo: efficient leave-one-out cross-validation and WAIC for Bayesian models}},
    author = {Vehtari, A. and Gabry, J. and Magnusson, M. and Yao, Y. and Bürkner, P.-C. and Paananen, T. and Gelman, A. and Goodrich, B. and Piironen, J. and Nicenboim, B. and Lindgren, L.},
    Year = {2024},
    Note = {{R package version 2.8.0}},
    Url = {https://cran.r-project.org/web/packages/loo/index.html}}

@article{Hamura03012025,
    author = {Hamura, Y. and Irie, K. and Sugasawa, S.},
    title = {{Robust Bayesian modeling of counts with zero inflation and outliers: theoretical robustness and efficient computation}},
    journal = {Journal of the American Statistical Association},
    doi = {https://doi.org/10.1080/01621459.2024.2447111},
    pages = {0--0},
    number = {0},
    volume = {0},
    year = {2025}}

@article{otto2025contaminated,
  title={{A contaminated regression model for count health data}},
  author={Otto, A. F. and Ferreira, J. T. and Tomarchio, S. D. and Bekker, A. and Punzo, A.},
  number={2},
  volume={34},
  pages={369--389},
  journal={Statistical Methods in Medical Research},
  doi={https://doi.org/10.1177/09622802241307613},
  year={2025}}

@article{karlis2005mixed,
  title={{Mixed Poisson distributions}},
  doi={https://doi.org/10.1111/j.1751-5823.2005.tb00250.x},
  author={Karlis, D. and Xekalaki, E.},
  journal={International Statistical Review},
  volume={73},
  number={1},
  pages={35--58},
  year={2005}}

@book{hilbe2011negative,
  title={Negative Binomial Regression},
  author={Hilbe, Joseph M},
  year={2011},
  publisher={Cambridge University Press}}

@article{nakagawa1975discrete,
  title={{The discrete Weibull distribution}},
  author={Nakagawa, T. and Osaki, S.},
  journal={IEEE Transactions on Reliability},
  volume={24},
  number={5},
  pages={300--301},
  doi={10.1109/TR.1975.5214915},
  year={1975}}

@phdthesis{kalktawi2017,
  author = {Kalktawi, H. S.},
  title = {{Discrete Weibull Regression Model for Count Data}},
  school = {Brunel University London},
  year = {2017},
  type = {{Ph.D. thesis}}}

@book{folland1999real,
  title={{Real Analysis: Modern Techniques and Their Applications}},
  author={Folland, G. B.},
  year={1999},
  publisher={John Wiley \& Sons}}

@book{rosen2019discrete,
  author    = {Rosen, K. H.},
  title     = {{Discrete Mathematics and Its Applications}},
  edition   = {8th},
  year      = {2019},
  publisher = {McGraw-Hill Education},
  address   = {New York, NY, USA}}

@misc{ArizonaMedpar1991,
  author       = "{National Health Economics \& Research Co.}",
  title        = "{1991 Arizona Medpar data, cardiovascular patient files}",
  howpublished = "{Arizona, U.S.}",
  year         = 1991}

@Manual{RCOUNT2025,
  title        = {{COUNT: functions, data and code for count data}},
  author       = {Hilbe, J. M. and Robinson, R.},
  year         = {2025},
  date         = {2025-01-20},
  note         = {{R package version 1.3.4}},
  url          = {https://CRAN.R-project.org/package=COUNT}}

@article{gelman1992inference,
  title={{Inference from iterative simulation using multiple sequences}},
  author={Gelman, A. and Rubin, D. B.},
  journal={Statistical Science},
  doi={https://doi.org/10.1214/ss/1177011136},
  volume={7},
  number={4},
  pages={457--472},
  year={1992}}

@book{Hampel2011,
  title     = {{Robust Statistics: The Approach Based on Influence Functions}},
  author    = {Hampel, F. R. and Ronchetti, E. M. and Rousseeuw, P. J. and Stahel, W. A.},
  publisher = {John Wiley \& Sons},
  series={Wiley Series in Probability and Statistics},
  year      = {2011}}

@article{Fihn2012,
  author    = {Fihn, S. D. and Gardin, J. M. and Abrams, J. and Berra, K. and Blankenship, J. C. and Dallas, A. P. and Douglas, P. S. and Foody, J. M. and Gerber, T. C. and Hinderliter, A. L. and King, S. B. and Kligfield, P. D. and Krumholz, H. M. and Kwong, R. Y. K. and Lim, M. J. and Linderbaum, J. A. and Mack, M. J. and Munger, M. A. and Prager, R. L. and Sabik, J. F. and Shaw, L. J. and Sikkema, J. D. and Smith, C. R. and Smith, S. C. and Spertus, J. A. and Williams, S. V.},
  title     = {2012 {ACCF}/{AHA}/{ACP}/{AATS}/{PCNA}/{SCAI}/{STS} guideline for the diagnosis and management of patients with stable ischemic heart disease: executive summary: a report of the {American College of Cardiology Foundation/American Heart Association Task Force on Practice Guidelines, and the American College of Physicians, American Association for Thoracic Surgery, Preventive Cardiovascular Nurses Association, Society for Cardiovascular Angiography and Interventions, and Society of Thoracic Surgeons}},
  journal   = {Journal of the American College of Cardiology},
  year      = {2012},
  volume    = {60},
  number    = {24},
  pages     = {2564--2603},
  doi       = {https://doi.org/10.1016/j.jacc.2012.07.012}}

@article{cantoni2001robust,
  title={{Robust inference for generalized linear models}},
  author={Cantoni, E. and Ronchetti, E.},
  doi={https://doi.org/10.1198/016214501753209004},
  journal={Journal of the American Statistical Association},
  volume={96},
  number={455},
  pages={1022--1030},
  year={2001}}

@article{beath2018mixture,
  title={{A mixture-based approach to robust analysis of generalised linear models}},
  author={Beath, K. J.},
  journal={Journal of Applied Statistics},
  doi={https://doi.org/10.1080/02664763.2017.1414164},
  volume={45},
  number={12},
  pages={2256--2268},
  year={2018}}

@article{datta2016bayesian,
  title={{Bayesian inference on quasi-sparse count data}},
  doi={https://doi.org/10.1093/biomet/asw053},
  author={Datta, J. and Dunson, D. B.},
  journal={Biometrika},
  volume={103},
  number={4},
  pages={971--983},
  year={2016}}

@article{SAMPFORD1955,
    author = {Sampford, M. R.},
    title = {The truncated negative binomial distribution},
    journal = {Biometrika},
    volume = {42},
    number = {1-2},
    pages = {58--69},
    year = {1955},
    doi = {10.1093/biomet/42.1-2.58}}

@article{hardin2015regression,
  title={{Regression models for count data from truncated distributions}},
  author={Hardin, J. W. and Hilbe, J. M.},
  journal={The Stata Journal},
  volume={15},
  number={1},
  pages={226--246},
  year={2015}}

@article{okhli2021contaminated,
  title={{On the contaminated exponential distribution: A theoretical Bayesian approach for modeling positive-valued insurance claim data with outliers}},
  author={Okhli, K. and Nooghabi, M. J.},
  journal={Applied Mathematics and Computation},
  volume={392},
  pages={125712},
  year={2021}
}

@article{otto2026modeling,
    title = {{Modeling bounded count environmental data using a contaminated beta-binomial regression model}},
    author = {Otto, A. F. and Punzo, A. and Ferreira, J. T. and Bekker, A. and Tomarchio, S. D. and Tortora, C.},
    journal = {Environmetrics},
    volume = {37},
    number = {1},
    pages = {e70067},
    year = {2026}
}

    \section*{Supporting information}

    Additional supporting information can be found online.

    \newpage
    \clearpage

    \setcounter{page}{1}
    \setcounter{figure}{0}
    \setcounter{table}{0}
    \renewcommand{\restoreapp}{}
    \renewcommand\appendixname{Web Appendix}
    \renewcommand{\figurename}{Web Figure}
    \renewcommand{\tablename}{Web Table}
    \renewcommand{\headrulewidth}{0pt}
    \titleformat{\section}{\large\bfseries}{\appendixname~\thesection .}{0.5em}{}
    
    \begin{appendices}
    
        {\Large\bf Robust median regression for count data with general lower truncation using a contaminated discrete Weibull model}
        
        \vskip 1.0cm
        
        {\normalsize Divan A. Burger, Janet van Niekerk, and Emmanuel Lesaffre}
        
        \vskip 4.5truecm
        
        \begin{center}
            \noindent
            {\Large\bf Supporting Information}
        \end{center}

        \newpage
        \clearpage
        
        \section{Truncated negative binomial distribution} \label{sec:TNB_DIST}
        
        Let $X \sim \mathrm{NB}\left(\mu, \alpha\right)$ with mean $\mu > 0$ and dispersion $\alpha > 0$. Its PMF is
        \begin{equation}
            P_{\mathrm{NB}}\left(x \left|\mu, \alpha\right.\right)
            = \frac{\Gamma\left(x + \alpha\right)}{\Gamma\left(\alpha\right)\Gamma\left(x + 1\right)}
            \left(\frac{\alpha}{\alpha + \mu}\right)^{\alpha}
            \left(\frac{\mu}{\alpha + \mu}\right)^{x},
            \quad x \in \left\{0,1,2,\ldots\right\}.
        \end{equation}
        Let $F_{\mathrm{NB}}\left(x \left|\mu, \alpha\right.\right) = \sum_{k=0}^{x} P_{\mathrm{NB}}\left(k \left|\mu, \alpha\right.\right)$ denote its CDF.
        
        Truncating at $c \in \left\{1,2,\ldots\right\}$ means defining
        \begin{equation}
            Y = X \left|\left(X \ge c\right)\right. .
        \end{equation}
        The TNB PMF and CDF are
        \begin{equation}
            P_{\mathrm{TNB}}\left(Y = y \left|\mu, \alpha, c\right.\right)
            = \frac{P_{\mathrm{NB}}\left(y \left|\mu, \alpha\right.\right)}
            {1 - F_{\mathrm{NB}}\left(c - 1 \left|\mu, \alpha\right.\right)},
            \qquad y \in \left\{c, c+1, \ldots\right\},
        \end{equation}
        \begin{equation}
            F_{\mathrm{TNB}}\left(y \left|\mu, \alpha, c\right.\right)
            = \frac{F_{\mathrm{NB}}\left(y \left|\mu, \alpha\right.\right) - F_{\mathrm{NB}}\left(c - 1 \left|\mu, \alpha\right.\right)}
            {1 - F_{\mathrm{NB}}\left(c - 1 \left|\mu, \alpha\right.\right)},
            \qquad y \ge c.
        \end{equation}
        For $c \in \left\{1,2,\ldots\right\}$, the truncated mean is
        \begin{equation}
            E\left(Y \left|\mu, \alpha, c\right.\right)
            = E\left(X \left|X \ge c\right.\right)
            = \frac{ \mu - \displaystyle\sum_{y=0}^{c-1} y P_{\mathrm{NB}}\left(y \left|\mu, \alpha\right.\right) }
            { 1 - F_{\mathrm{NB}}\left(c - 1 \left|\mu, \alpha\right.\right) }.
        \end{equation}
        Substituting the NB PMF gives an explicit form for the lower-tail first moment:
        \begin{equation}
            \sum_{y=0}^{c-1} y P_{\mathrm{NB}}\left(y \left|\mu, \alpha\right.\right)
            = \sum_{y=1}^{c-1}
            y \frac{\Gamma\left(y + \alpha\right)}{\Gamma\left(\alpha\right)\Gamma\left(y + 1\right)}
            \left(\frac{\alpha}{\alpha + \mu}\right)^{\alpha}
            \left(\frac{\mu}{\alpha + \mu}\right)^{y},
        \end{equation}
        and, using $y\Gamma\left(y+\alpha\right)/\Gamma\left(y+1\right)=\Gamma\left(y+\alpha\right)/\Gamma\left(y\right)$,
        \begin{equation}
            \sum_{y=0}^{c-1} y P_{\mathrm{NB}}\left(y \left|\mu, \alpha\right.\right)
            = \left(\frac{\alpha}{\alpha + \mu}\right)^{\alpha}
              \left(\frac{\mu}{\alpha + \mu}\right)
              \sum_{k=0}^{c-2}
              \frac{\Gamma\left(k + \alpha + 1\right)}{\Gamma\left(\alpha\right)\Gamma\left(k + 1\right)}
              \left(\frac{\mu}{\alpha + \mu}\right)^{k}.
        \end{equation}

        \newpage

        \section{Figures} \label{sec:WEB_FIGURES}

        \begin{figure}[h]
            \centering
            \includegraphics[width=0.7\textwidth]{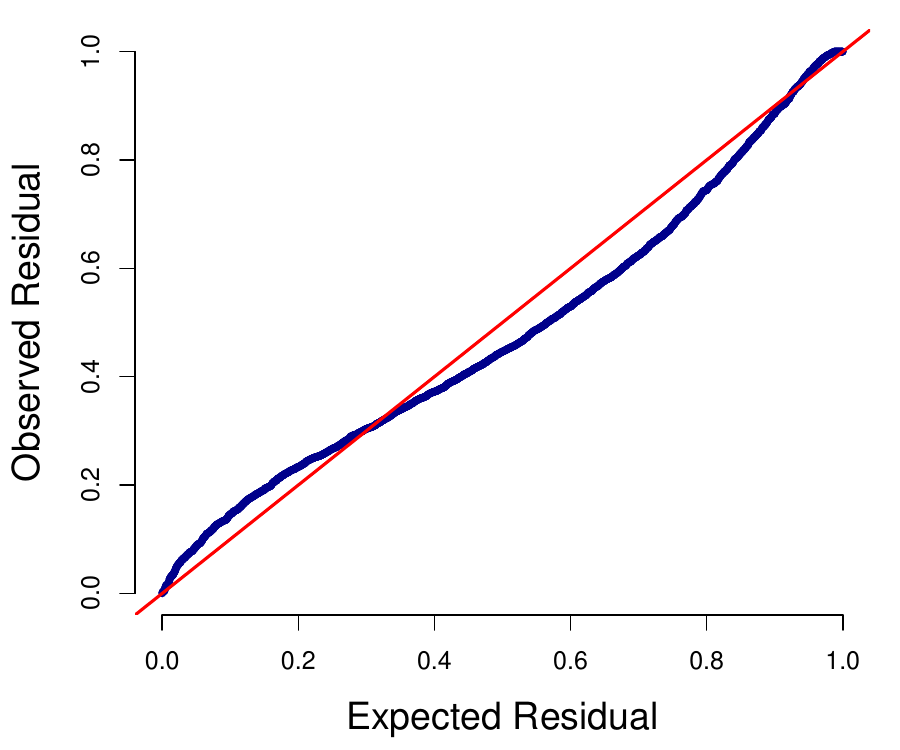}
            \caption{QQ plot of simulation-based residuals for the TNB model. The blue points show the empirical distribution of residuals against the ideal uniform distribution (red diagonal). Departures from the diagonal mirror those under the single-component TDW, indicating a comparable lack of fit.}
            \label{fig:LOS_TNB_residuals}
        \end{figure}

        \begin{figure}[t]
            \centering
            \includegraphics[width=0.75\textwidth]{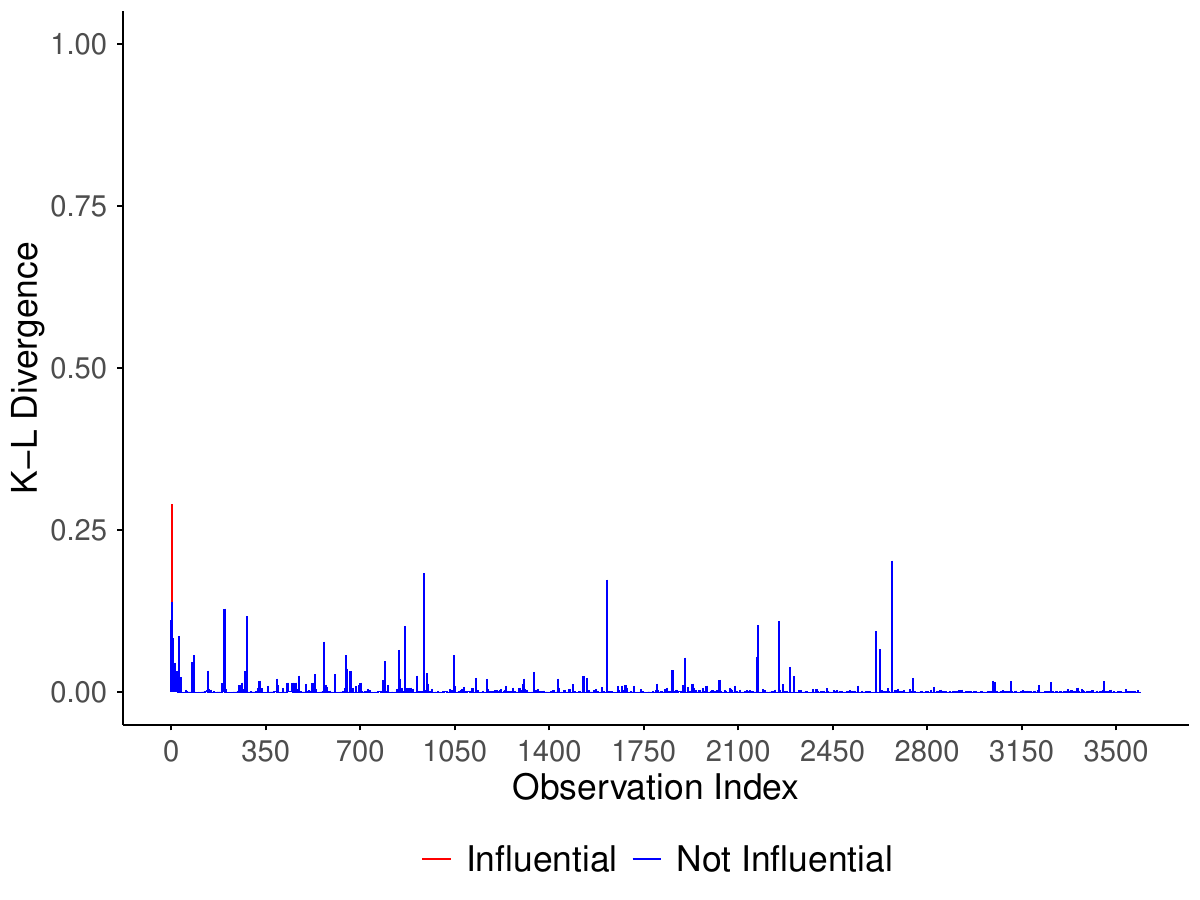}
            \caption{K-L divergence plot for the TNB model. Each vertical bar represents an observation, colored red if classified as influential and blue otherwise. Compared with the single-component TDW, the TNB exhibits fewer and smaller influential points, indicating a more stable influence profile.}
            \label{fig:LOS_TNB_KL}
        \end{figure}

        \newpage
        \clearpage

        \begin{landscape}
        \pagestyle{landscape}
            \begin{figure}[htbp]
                \centering
                \includegraphics[scale = 0.8]{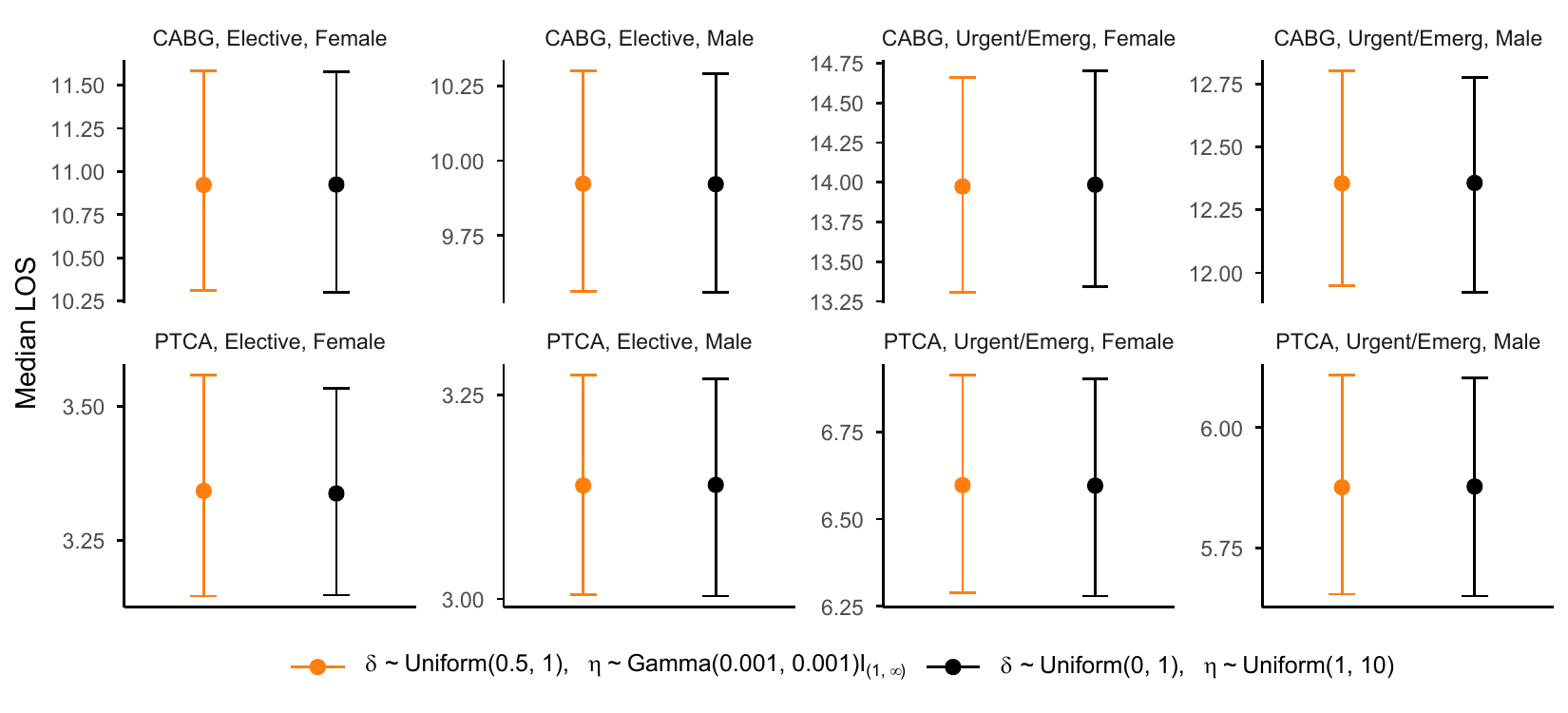}
                \caption{Posterior estimates of the shifted median $m^*$ for each subgroup under the cTDW model with two prior specifications. Each point is the posterior median of $m^*$, and the vertical bars show the 95\% BCI. Facets distinguish the combinations of procedure type, admission category, and patient sex. Orange corresponds to the main prior specification, $\left(\delta, \eta\right) \sim \left(\mathrm{Uniform}\left(0.5,1\right), \mathrm{Gamma}\left(0.001, 0.001\right)\mathbb{I}_{\left(1,\infty\right)}\right)$, whereas black corresponds to the alternative prior specification used in the sensitivity analysis, $\left(\delta, \eta\right) \sim \left(\mathrm{Uniform}\left(0,1\right), \mathrm{Uniform}\left(1,10\right)\right)$. Posterior median estimates and 95\% BCIs are very similar under both specifications.}
                \label{fig:median_bands_cDW}
            \end{figure}
        \end{landscape}

        \begin{figure}[ht]
            \centering
            \includegraphics[width=0.64\textwidth]{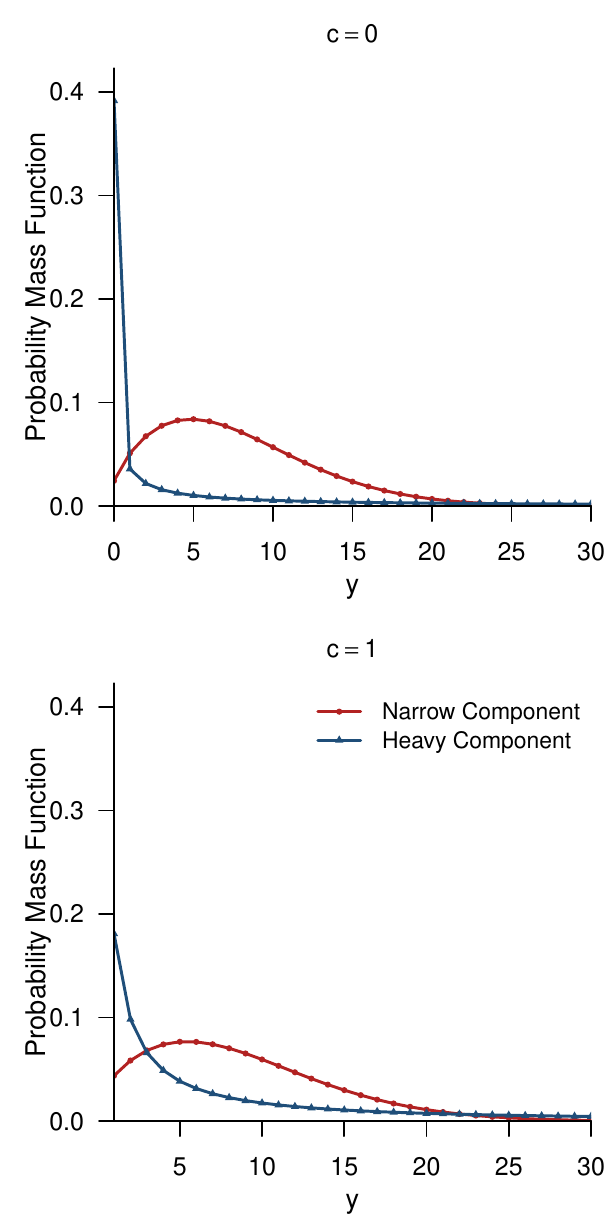}
            \caption{Separation between the narrower and heavier-tail TDW components under $c = 0$ (top panel) and $c = 1$ (bottom panel), with the linear predictor held fixed at $\beta_0 = 2$, so that $m^* = \exp\left(2\right)$ for $c = 0$ and $m^* = 1 + \exp\left(2\right)$ for $c = 1$. The red curve denotes the narrower component, and the blue curve denotes the heavier-tail component. When $c = 0$, the zero cell contributes lower-tail information that helps distinguish the two components. After truncation at $c = 1$, this information is no longer available, which provides a plausible explanation for the weaker finite-sample identification of $\alpha$, $\eta$, and $\delta$ observed in the simulation study.}
            \label{fig:ComponentSeparation}
        \end{figure}

        \newpage
        \clearpage

        \section{Tables} \label{sec:WEB_TABLES}

        \begin{longtable}{
            l 
            S[table-format=3.3] 
            S[table-format=3.3] 
            S[table-format=2.3] 
            l 
            S[table-format=3.3] 
            S[table-format=3.3] 
            S[table-format=2.3]}
            \caption{Posterior medians (Est.) and 95\% BCIs for the cTDW model applied to the Arizona hospital LOS data under two prior specifications. The regression coefficients $\beta_j$ link covariates to a shifted median via $\log\left(m_i^* - 1\right) = \mathbf{x}_i^\top \bm{\beta}$, where $\beta_1$ through $\beta_7$ capture the effects of procedure type (CABG vs.\ PTCA), admission category (urgent/emergent vs.\ elective), and sex (male vs.\ female), including interactions. The dispersion parameter is $\alpha$, while $\eta$ inflates the heavier-tail subdistribution and $\delta$ denotes the mixture weight of the narrower component, so that the heavier-tail component has weight $1 - \delta$. The main prior specification is $\left(\delta, \eta\right) \sim \left(\mathrm{Uniform}\left(0.5, 1\right), \mathrm{Gamma}\left(0.001, 0.001\right)\mathbb{I}_{\left(1,\infty\right)}\right)$, whereas the alternative prior specification used in the sensitivity analysis is $\left(\delta, \eta\right) \sim \left(\mathrm{Uniform}\left(0, 1\right), \mathrm{Uniform}\left(1, 10\right)\right)$.} \label{tab:median_bands_cDW} \\
                
            \hline
            & \multicolumn{7}{c}{\textbf{cTDW Model}} \\
            \cline{2-8}
            & \multicolumn{3}{c}{\textbf{Main Prior}} & & \multicolumn{3}{c}{\textbf{Alternative Prior}} \\
            \cline{2-4} \cline{6-8}
            \textbf{Parameter} & \textbf{Est.} & \multicolumn{2}{c}{\textbf{95\% BCI}} & & \textbf{Est.} & \multicolumn{2}{c}{\textbf{95\% BCI}} \\
            \hline
            \endfirsthead
                
            & \multicolumn{7}{c}{\textbf{cTDW Model}} \\
            \cline{2-8}
            & \multicolumn{3}{c}{\textbf{Main Prior}} & & \multicolumn{3}{c}{\textbf{Alternative Prior}} \\
            \cline{2-4} \cline{6-8}
            \textbf{Parameter} & \textbf{Est.} & \multicolumn{2}{c}{\textbf{95\% BCI}} & & \textbf{Est.} & \multicolumn{2}{c}{\textbf{95\% BCI}} \\
            \hline
            \endhead
                
            \hline
            \multicolumn{8}{r}{{Continued on next page}}\\
            \endfoot
                
            \hline
            \endlastfoot
                
              $\beta_{0}$ & 0.851 & [0.765; & 0.942] &  & 0.849 & [0.766; & 0.933] \\ 
              $\beta_{1}$ & 1.444 & [1.333; & 1.553] &  & 1.446 & [1.343; & 1.549] \\ 
              $\beta_{2}$ & 0.871 & [0.767; & 0.974] &  & 0.873 & [0.776; & 0.973] \\ 
              $\beta_{3}$ & -0.090 & [-0.198; & 0.015] &  & -0.088 & [-0.188; & 0.012] \\ 
              $\beta_{4}$ & -0.604 & [-0.734; & -0.470] &  & -0.603 & [-0.731; & -0.482] \\ 
              $\beta_{5}$ & -0.016 & [-0.144; & 0.116] &  & -0.019 & [-0.143; & 0.109] \\ 
              $\beta_{6}$ & -0.047 & [-0.176; & 0.080] &  & -0.049 & [-0.174; & 0.074] \\ 
              $\beta_{7}$ & 0.020 & [-0.140; & 0.180] &  & 0.022 & [-0.133; & 0.177] \\ 
              $\alpha$ & 0.303 & [0.280; & 0.326] &  & 0.303 & [0.280; & 0.326] \\ 
              $\eta$ & 2.846 & [2.655; & 3.053] &  & 2.848 & [2.657; & 3.053] \\ 
              $\delta$ & 0.688 & [0.631; & 0.741] &  & 0.688 & [0.632; & 0.741] \\ 
            \hline
        \end{longtable}

    \end{appendices}

\end{document}